\def\p{\phantom{1}}

\def\lae{\mathrel{<\kern-1.0em\lower0.9ex\hbox{$\sim$}}}
\def\gae{\mathrel{>\kern-1.0em\lower0.9ex\hbox{$\sim$}}}

\documentclass{emulateapj}
\usepackage{longtable}

\slugcomment{Accepted for publication in ApJ}
\lefthead{{JORD\'AN ET AL.}}
\righthead{LOW MASS X-RAY BINARIES AND GLOBULAR CLUSTERS IN M87}

\begin{document}

\title{The ACS Virgo Cluster Survey III. {\it Chandra} and {\it HST}
Observations of Low-Mass X-Ray Binaries and Globular Clusters in
M87\altaffilmark{1}}

\author{Andr\'es Jord\'an\altaffilmark{2,3}, Patrick C\^ot\'e\altaffilmark{2},
Laura Ferrarese\altaffilmark{2}, John P. Blakeslee\altaffilmark{4}, 
Simona Mei\altaffilmark{5}, David Merritt\altaffilmark{2}, 
Milo\v{s} Milosavljevi\'c\altaffilmark{6}, 
Eric W. Peng\altaffilmark{2},
John L. Tonry\altaffilmark{7} and
Michael J. West\altaffilmark{8}}

\begin{abstract}
The ACIS instrument on board the {\it Chandra} X-ray Observatory has been used to carry
out the first systematic study of low-mass X-ray binaries (LMXBs) in M87, the
giant elliptical galaxy near the dynamical center of the Virgo Cluster. 
These images --- having a total exposure time of 154~ks --- are the deepest
X-ray observations yet obtained of M87. We identify 174 X-ray point-sources,
of which $\sim$ 150 are likely LMXBs. This LMXB catalog is combined with
deep F475W and F850LP images taken with ACS on {\it HST} (as part of the ACS
Virgo Cluster Survey) to examine the connection between LMXBs and globular clusters
in M87. Of the 1688 globular clusters in our catalog, $f_X = 3.6\pm$0.5\%
contain a LMXB. Dividing the globular cluster sample by 
metallicity, we find that the metal-rich clusters are 3$\pm$1 times more
likely to harbor a LMXB than their metal-poor counterparts. In agreement
with previous findings for other galaxies based on smaller LMXB samples,
we find the efficiency of LMXB formation to scale with both cluster 
metallicity, $Z$, and luminosity, in the sense that brighter, more metal-rich
clusters are more likely to contain a LMXB. For the first time, however, we
are able to demonstrate that the probability, $p_X$, that a given cluster will contain
a LMXB depends sensitively on the {\it dynamical properties} of the host
cluster. Specifically, we use the HST images to measure the half-light radius,
concentration index and central density, $\rho_0$, for each globular, and define
a parameter, $\Gamma$, which is related to the tidal capture and
binary-neutron star exchange rate. Our preferred form for $p_X$ is then
$p_{X}\propto \Gamma\rho_0^{-0.42\pm0.11}(Z/Z_{\odot})^{0.33\pm0.1}.$
We argue that if the form of $p_X$ is determined by dynamical processes, 
then the observed metallicity dependence is a consequence of 
an increased number of neutron stars per unit mass in metal-rich 
globular clusters.
Finally, we present a critical examination of the LMXB luminosity function 
in M87 and re-examine the published LMXB luminosity functions for M49 and NGC~4697.
We find no compelling evidence for a break in the luminosity distribution
of resolved X-ray point sources in any of these galaxies. Instead, the
LMXB luminosity function in all three galaxies is well described by a
power law with an upper cutoff at $L_X \sim 10^{39}$~erg~s$^{-1}$.
\end{abstract}


\altaffiltext{1}{Based on observations with the NASA/ESA 
{\it Hubble Space Telescope}
obtained at the Space Telescope Science Institute, which is operated 
by the Association
of Universities for Research in Astronomy, Inc., under 
NASA contract NAS 5-26555}
\altaffiltext{2}{Department of Physics and Astronomy, Rutgers University, 
Piscataway, NJ 08854;
andresj@physics.rutgers.edu, pcote@physics.rutgers.edu, 
lff@physics.rutgers.edu, 
merritt@physics.rutgers.edu, ericpeng@physics.rutgers.edu}
\altaffiltext{3}{Claudio Anguita Fellow}
\altaffiltext{4}{Department of Physics and Astronomy, 
Johns Hopkins University, Baltimore, MD 21218; jpb@pha.jhu.edu.}
\altaffiltext{5}{Institut d'Astrophysique Spatiale, 
Universit\'e Paris-Sud, B\^at. 121, 
91405 Orsay, France; simona.mei@ias.u-psud.fr.}
\altaffiltext{6}{Theoretical Astrophysics, California Institute of Technology,
Pasadena, CA 91125; milos@tapir.caltech.edu.}
\altaffiltext{7}{Institute of Astronomy, University of Hawaii, 
2680 Woodlawn Drive, Honolulu, HI 96822; jt@ifa.hawaii.edu}
\altaffiltext{8}{Department of Physics \& Astronomy, University of Hawaii, 
Hilo, HI 96720; west@astro.uhh.hawaii.edu.}

\section{Introduction}
\label{sec:intro}

Normal elliptical galaxies have long been known to harbor two major 
components of X-ray emission --- a soft component due to emission from
diffuse gas, and a harder one arising from a population of 
low-mass X-ray binaries (LMXBs). The existence of the latter component 
was first inferred  from the spectral
hardening of elliptical galaxies with progressively smaller
X-ray to optical luminosities, a trend reminiscent of
late-type galaxies in which a portion of the X-ray emission could be
identified directly with a population of accreting binary stars 
(Kim, Fabbiano \& Trinchieri 1992). With the launch of \textit{Chandra},
the harder X-Ray component has been partially resolved into
point sources associated with LMXBs
for an ever-increasing number of early-type
galaxies ($e.g.$, Sarazin, Irwin \& Bregman 2001, 2002; 
Angelini, Loewenstein \& Mushotzky 2001; Blanton, Sarazin
\& Irwin 2001; Finoguenov \& Jones 2002; Kundu, Maccarone
\& Zepf 2002; Maccarone, Kundu \& Zepf 2003; Kundu et~al. 2003; Jeltema et~al. 2003;
Kim \& Fabbiano 2003;
Irwin, Athey \& Bregman 2003; Sivakoff, Sarazin \& Irwin 2003). 

In the Milky Way, it was realized soon after 
the discovery of X-ray emission from a handful
of Galactic globular clusters (GCs) (Giacconi et~al. 1974; Forman
et~al. 1978) that the number of X-ray sources
per unit mass is several hundred times higher in GCs than
in the halo field (Katz 1975; Clark 1975). 
The population of X-ray point sources associated with GCs was later found
to be a mixture of dim ($L_X \lae 10^{34.5}$~erg~s$^{-1}$) and
bright ($10^{36} \lae L_X \lae 10^{38}$~erg~s$^{-1}$) 
sources (Hertz \& Grindlay 1983). The fainter sources seem to be
composed of many different kinds of objects (see, $e.g.$, Verbunt 2001),
while the bright sources are believed to be accreting neutron stars, and
are thus classified as LMXBs.
Since GCs contain only $\lae 0.1$\%
of the Galaxy's stars, but $\sim 10\%$ of its LMXBs, it is clear
that GCs are efficient sites of LMXB formation. According to current
thinking, the overabundance of LMXBs in GCs is a direct consequence
of their high central densities. In such environments, the rates
of tidal capture of neutron stars and single-binary exchange interactions
--- the two principal mechanisms by which LMXB progenitors are thought
to be produced --- are greatly enhanced relative to the field
(Clark 1975; Fabian, Pringle \& Rees 1975; Hills 1976).

{\it Chandra}'s ability to resolve LMXBs in nearby galaxies allows
us to examine the connection between LMXBs and GCs in new and different
environments. 
Such studies may yield valuable information on the 
processes by which LMXBs form and evolve. 
Moreover, what is lost in the detailed
description of individual LMXBs and GCs beyond the Local Group
is counter-balanced by the potentially dramatic gains in sample
size (for example, the total number of bright X-ray sources
belonging to the Galactic GC system is limited to just thirteen objects; 
Verbunt 2001, White \& Angelini 2001).  M87 (NGC~4486), the giant elliptical galaxy near
the dynamical center of the Virgo Cluster, has the richest GC system 
in the local supercluster with a total of 13,450$\pm$950 GCs 
(McLaughlin, Harris \& Hanes 1994). 
It is also one of the most thoroughly studied
GC systems, with a wealth of spectroscopic and photometric information 
available on its metallicity distribution, spatial structure,
luminosity function, age distribution, and dynamical properties
(McLaughlin, Harris \& Hanes 1994; Cohen, Blakeslee \& Ryzhov 1998;
Harris et~al. 1998; Kundu et~al. 1999; Hanes et~al. 2001; C\^ot\'e et~al. 2001;
Kissler-Patig, Brodie \& Minniti 2002; Jord\'an et~al. 2002). 
Yet nothing is known about the LMXB population in M87 or its relation
to the M87 GC system, partly due to the difficulty of detecting 
individual X-ray point sources superimposed on a bright, and spatially
varying
diffuse background.
In this paper, we combine deep X-ray observations from \textit{Chandra}
with optical  $g_{475}$ and $z_{850}$ imaging from {\it HST} to characterize the
LMXB population in M87, and to examine its connection to the 
underlying GC system.

In what follows, we assume a distance to M87 of $D = 16$ Mpc (Tonry et~al. 2001),
an effective radius of $R_e = 96''$ (de Vaucouleurs \& Nieto 1978), and 
a Galactic column density of $N_H=2.5\times 10^{20}$~cm$^{-2}$ (Stark et~al. 1992).
Before starting, we set some notation and conventions.
We will make repeated use of the
Kolmogorov-Smirnov (KS) test and the Wilcoxon rank sum test
(Wilcoxon 1945; Mann \& Whitney 1947). The KS test tests the hypothesis
that two samples are drawn from the same parent
continuous distribution
(two-sample KS test) or that a given sample was drawn
from a specified continuous distribution (one-sample KS test), whereas the
Wilcoxon rank sum tests the hypothesis that the
location of the samples' parent distributions is the same.
The results of these tests will be often reported by their
returned p-values, which give the probability of
obtaining the observed statistic under the null hypothesis.
When talking about probability density functions the
symbol $\sim$ should be understood in the sense of ``distributes
as'' rather than its usual sense, and
we will not write explicitly in those distributions
the normalization constants.

\section{Observations and Data Reductions}
\label{sec:red}

\subsection{X-ray catalog}
\label{sec:xraycat}

M87 (=NGC~4486) was observed with the \textit{Chandra}
Advanced CCD Imaging Spectrometer ({\it ACIS})
for $121$ ks on 5--6 July 2002. In what follows, we use
only the S3 chip data. The data were processed following 
the CIAO data reduction threads, including a correction for charge 
transfer inefficiency (CTI; Townsley et~al. 2000). 
Additionally, we used $38$ ks of archival {\it ACIS} observations 
of M87 taken on 29 July 2000
(PI: A.S. Wilson). These data were processed in similar 
fashion to the July 2002 data, except that no CTI 
correction was possible because the data were telemetered in Graded mode.
All reductions were carried out with CIAO $2.3$ coupled with CALDB $2.21$. 
In order to combine the events files into a single image for 
point source detection, we obtained relative offsets by matching 
the celestial coordinates of two X-ray point 
sources\footnote{CXOU J123047.1+122415 and CXOU 123044.6+122140}. The 
relative offset was $\approx 0\farcs 5$. After excluding the
provisional list of point
sources identified with SExtractor (Bertin \& Arnouts 1996),
the M87 jet, and the central regions of the galaxy,
we created a light curve binned in $50$~s time intervals in
order to clean background flares by rejecting
points deviating by more than $3\sigma$ from the quiescent mean. 
The total exposure time of the co-added image, excluding four 
flares totaling $\approx$2.5 ksec,
was 154 ks.

Detection was performed using a two stage process. 
First, a wavelet filtering of the image 
was performed to keep only those objects with a characteristic
structure $\lae 8$ pixels and 
a significance greater than $2\times10^{-5}$ ($\sim 4\sigma$) using the
MR/1 package (Starck, Murtagh \& Bijaoui 1998). 
Object detection was then performed on this
filtered image using SExtractor.
Note that this detection procedure was also used by 
Valtchanov, Pierre \& Gastaud (2001)
who found it to be the most effective  for {\it XMM-Newton}
data. The inner regions of M87 exhibit
a wealth of structure in its diffuse emission, with numerous
bubbles and arcs that cause spurious point source detections. 
These regions, along with the jet and the central source, were masked. 
Source regions were defined to be ellipses
with semi-major and semi-minor axes equal to twice the parameter values
returned by SExtractor. We also used \textit{wavdetect} (Freeman et~al. 2002)
to carry out the source detection, producing a similar point source
catalog in the process, but a visual inspection of the
individual detections suggested that the adopted method produced
superior determinations of the source centroids
(which are of critical importance when matching to the optical sources). 
Indeed, using positions obtained with \textit{wavdetect} the rms of the 
difference in celestial coordinates between the X-ray and optical sources
increased by $\sim 35\%$.

The total number of detected point sources in the S3 chip was $174$, 
of which $\sim 20$ are expected to be background contaminants such as AGN (Mushotzky et~al. 2000; 
Giacconi et~al. 2001).  A background region
was defined for each source by taking an annulus centered on the
source. Since the background varies on small spatial scales in the inner
regions of the galaxy, the inner radii of the annuli were taken to be
\begin{equation}
\begin{array}{rclrr}
r_i & = & a_{maj} & {\rm for} & r \leq 0.44R_e \\
r_i & = & 1.5a_{maj} & {\rm for} & r > 0.44R_e \\
\end{array}{}
\label{eq1}
\end{equation}
where $a_{maj}$ is the semi-major axis of the source extraction region. For
all objects, the outer radius of the background annulus was calculated so that 
the total background area was five times that of the source extraction region.
For a few objects, the background annulus defined in this way included a second
point source. In such cases, the annulus was masked over the angular region 
containing the adjacent source, and the outer radius adjusted so that the area
of the background region was five times that of the source extraction region.

\subsection{Optical catalog}
\label{sec:optcat}

M87 was observed as part of the ACS Virgo Cluster Survey
(C\^ot\'e et~al. 2004, in preparation) on 19 January 2003. 
The full dataset consists of
two $375$ s exposures in the F475W ($g_{475}$) band, two $560$ s exposures
in the F850LP ($z_{850}$) band, and a single 90 s F850LP exposure.
A detailed account of the reduction procedures for the survey
will be presented elsewhere (Jord\'an et~al. 2004, in preparation) so we give
only a brief summary here. 

After determining the small shifts between the images, 
they were combined using the PYRAF task \textit{multidrizzle} 
(Koekemoer et~al. 2002). 
Detection images are built by subtracting a model of the galaxy and  
object detection on these images was performed independently for each band using SExtractor
(Bertin \& Arnouts 1996) with a threshold of $5$ connected pixels
at a level of $1.5\,\sigma$. The celestial coordinates of the detections were 
matched with a $0\farcs1$ matching radius; sources detected in just a
single filter were discarded.

GCs at the distance of M87 are slightly resolved with ACS. This
opens the possibility of modeling directly the two-dimensional
light distribution of the GCs. A code has been developed (Jord\'an \&
C\^ot\'e 2004, in preparation) to measure
total magnitude, half-light radius, $r_h$, and concentration index, $c$,
for each GC by fitting the two-dimensional ACS surface brightness
profiles with the convolution of the instrumental point spread
function (PSF) with isotropic, single-mass King (1966)
models. These models are well known to provide an excellent 
representation of the surface brightness profiles of most Galactic GCs.
DAOPHOT II (Stetson 1987; 1993) was used to derive PSFs that varied
quadratically with CCD position, using
archival observations of moderately crowded fields
in the outskirts of the Galactic GC $47$ Tucanae.
Instrumental magnitudes were converted to the AB system
using zeropoints of
$26.07$ mag for F475W and $24.86$ mag for F850LP (Sirianni et al. 2004).
A correction for foreground extinction was performed using the 
reddening curves of Cardelli, Clayton \& Mathis (1989), with a 
value of E(B-V) = 0.023 taken from the DIRBE maps 
of Schlegel, Finkbeiner \& Davis (1998).
Hereafter, we denote the F475W and F850LP filters by the corresponding
filters ($g_{475}$ and $z_{850}$, respectively) in the Sloan Digital Sky Survey.

In order to define a set of bonafide GCs using this optical
catalog, several additional selection criteria were imposed:
(1) a color in the range $0.4 < (g_{475}-z_{850}) < 1.9$, as measured from
both SExtractor and the King model fitting program; (2) a half-light
radius in the range $0.5 < r_h \mbox{(pc)} < 15.5 $, an interval that
encompasses almost all GCs in the Milky Way (Harris 1996); and (3) 
agreement between the $r_h$ measurements in the $g_{475}$ and $z_{850}$
bandpasses at the $4\sigma$ level.
Additionally, we removed from the optical catalog 106 GC candidates
which fall in the regions that were masked in the X-ray image 
(see \S~\ref{sec:xraycat}).
A total of 1688 GC candidates, spanning nearly six magnitudes in
brightness, met these selection criteria. When matching
to the X-ray point-source catalog, the full
list of optical point sources has been used, since it is likely
that some fraction of the X-ray point sources will be unrelated
to GCs.

\subsection{Matching}
\label{sec:matching}

The high density of GCs in M87 requires care to be taken in
matching the X-ray and optical catalogs. As an additional complication,
the factor of ten difference in pixel size between ACIS and ACS
makes any uncertainty in the X-ray coordinates translate into many
ACS pixels. 

The quality of the absolute pointing was first verified by comparing the
coordinates of the galaxy nucleus, in both the X-ray and optical images,
with the VLBI position of the M87 nucleus obtained
by Ma et~al. (1998):
$\alpha$(J2000)$=12^h30^m49.^s423381$ and 
$\delta$(J2000)$=12^{\circ}23'28.''0434$.  For both
the optical and X-ray images, the coordinates of
the brightest nucleus pixel differ by less than $\sim 0\farcs3$ from
the VLBI coordinates. We are thus confident that both datasets have
good absolute pointing.
An initial matching without any adjustment between 
the X-ray and optical sources
further confirmed the compatibility of the coordinates, and 
an analysis of the residuals 
showed no statistically significant trends as a function 
of position in the chip.

Given that any offset is small and there is no appreciable rotation 
between the two coordinate systems, we adopted the following 
iterative scheme to obtain constant offsets in each coordinate,
${\Delta}\alpha$ and ${\Delta}\delta$.
First, all sources within a radius of $0\farcs3$ were first matched,
and a biweight (Beers, Flynn \& Gebhardt 1990) estimate of the offsets
calculated. In the next iteration, the matching radius was then 
incremented to $0\farcs5$, improved biweight estimates of the offsets calculated, 
and the list of matched objects updated accordingly. This process was
repeated until the list of matched objects stabilized.

\begin{figure}
\plotone{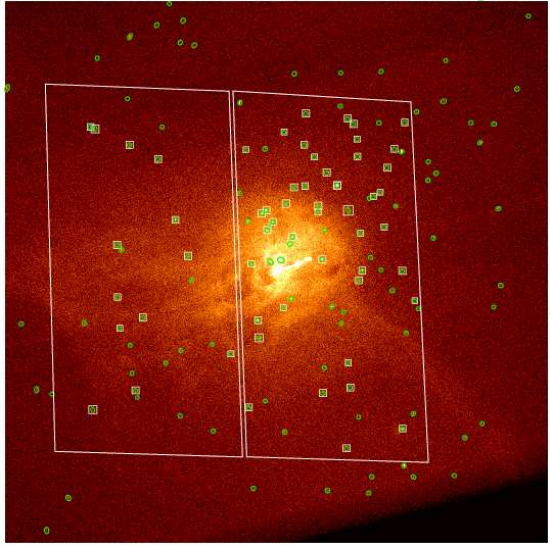}
\caption[ximg]{Co-added \textit{Chandra/ACIS} image of M87 with the
ACS field of view overlaid (rhomboids). The X-ray point source
detections  are indicated by the green ellipses. White squares indicate the
60 X-ray sources that coincide with globular cluster candidates.
North is up and East is to the left in this image, which
measures $5'\times5'$.
\label{fig:ximg}}
\end{figure}

The adopted offsets (in the sense optical minus X-ray) are
${\Delta}\alpha = -0\farcs1$ and ${\Delta}\delta = +0\farcs15$.
The rms deviation of the residuals are $0\farcs14$ and $0\farcs13$,
respectively. Figure~\ref{fig:ximg} shows the central $5'\times5'$
of the \textit{Chandra} image with the ACS field of view overlaid.
X-ray point sources are indicated by green ellipses; those
sources that coincide with optical GC candidates are 
marked with white squares. The final list contains $62$ optical sources 
of any sort that are matched to an X-ray source; $60$ of these optical sources are
probable GCs based on the criteria described in \S\ref{sec:optcat}.
Data for all X-ray point sources 
are given in Table~\ref{tab:cat}, which records the source identification
number, coordinates, count rate, luminosity and hardness ratios (see below). 
Column 7 gives a flag to indicate whether the X-ray source falls
within the ACS field of view, while comments
on the classification of the various optical sources are given in the
final column.

\begin{figure}
\plottwo{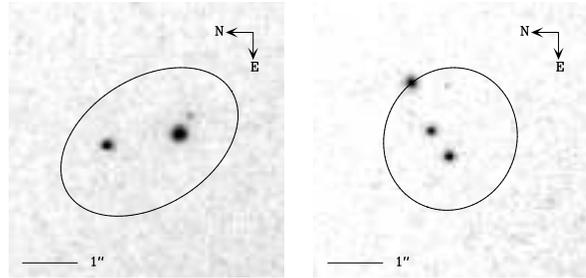}{f2b.eps}
\caption[nuc]{(Left Panel) Portion of the ACS F850LP image in which the 
extraction region for the X-ray source CXOU J123047.1+122415
is marked by an ellipse. The X-ray source contains three GC candidates,
none of which is within a matching radius of the 
X-ray emission centroid.
(Right Panel) Portion of the ACS F850LP image in which the
extraction region for the X-ray source CXOU 123046.7+122402 is marked by an ellipse.
The X-ray source contains two GC candidates within
one matching radius of its centroid.
\label{fig:zoom}
}
\end{figure}

Two sources in particular merit attention (see Figure~\ref{fig:zoom}). The extraction ellipse
of one X-ray source, CXOU J123047.1+122415 in Table~\ref{tab:cat}, encloses
three optical sources. In X-rays, the source is 
extended in a way that is consistent with being a blend of multiple sources.
The centroid of the X-ray emission lies approximately halfway between the
brightest optical detections, yet the three optical sources lie outside the nominal matching
radius and so do not make it into the final catalog. We consider this to
be a match when calculating the fraction of X-ray-optical matches by 
assuming that two of the GC matches hold an LMXB, but
discard this source from the subsequent analysis.
A second X-ray source (object CXOU J123046.7+122402 in Table~\ref{tab:cat}) has two optical
candidates within the matching radius. Given this ambiguity, we consider this
to be a match for the purposes of estimating the overall frequency of 
LMXB-GC associations, but do not include this source in any other aspect
of the analysis. All the candidate GCs in these two sources are metal-rich, so
no ambiguity is introduced when calculating the frequencies for the 
metal-rich and metal-poor groups. Removing the two optical matches
to CXOU J123046.7+122402 leaves 58 X-ray point source matches
that will be used in the analysis.

Given the high GC surface density in M87, it is of interest to know the number
of chance matches that might occur between the X-ray and optical datasets.
We have estimated the number of such false matches by rotating
the X-ray source coordinates about galaxy's nucleus and calculating the
total number of matches at each rotation angle. This exercise produced
an average of four matches at each angle, so we conclude that the
number of chance associations in our sample is small. According to the
background source counts of Giacconi et~al. (2001) and Mushotzky
et~al. (2000), we expect $\sim 2$ 
background sources within the
ACS field of view. This is comparable to the number (2) of
X-ray sources in our sample that match an optical source that is not
a probable GC,
so we believe that our sample has very little contamination from
background sources and false matches.

\subsection{Variability}

We performed a simple search for variability among the 23
X-ray sources with $L_X > 10^{38}$ erg s$^{-1}$
that coincide with a GC candidate. Using the positions from
the combined observations, we measured the fluxes for the 2000
and 2002 datasets. The distribution of the flux differences,
when divided by the expected uncertainty, reveals that seven
sources show significant variability when compared with a
normal distribution. Thus, the incidence of X-ray transients
in the M87 GC/LMXB population appears roughly consistent
with that of the Galaxy, where roughly half of all sources
are known to exhibit time variability (e.g., Verbunt et al. 1995).

\section{Spectral Analysis}

\begin{figure}
\includegraphics[angle=-90,scale=0.35]{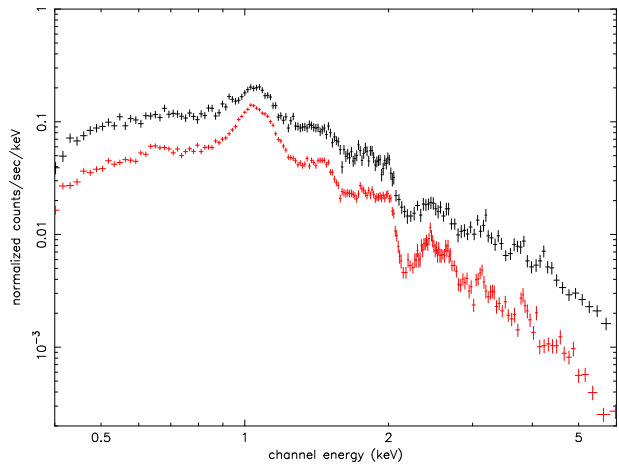}
\caption[xback_sources]{Summed spectra for the background 
regions (red crosses) and for source regions prior to background 
subtraction (black crosses). The spectra are based entirely on
data acquired in 2002.
\label{fig:xback_sources}
}
\end{figure}

All X-ray sources with galactocentric radii in the range $0.44R_e < r < 2R_e$,
were extracted and summed into a composite spectrum with the CIAO
routine \textit{acisspec}, which computes a weighted
redistribution matrix file and ancillary response file (ARF) 
appropriate for point sources distributed over the {\it ACIS} detector.  
The ARF files were corrected for the degradation of the {\it ACIS} 
quantum efficiency using the CIAO tool \textit{apply\_acisabs},
which applies the ACISABS absorption 
profile (Chartas \& Getman 2002)\footnote{See
\url{http://www.astro.psu.edu/users/chartas/xcontdir/xcont.html}}
to the original ARF file. The extraction was performed independently
for the 2000 and 2002 datasets, and the source 
spectra were regrouped so that each energy channel contained a
minimum of $50$ photons, prior to background subtraction. 
In Figure~\ref{fig:xback_sources}, we show summed spectra from 
the 2002 dataset, for both the source regions ($i.e.,$ object plus
background) and for the background regions alone.

\begin{figure}
\includegraphics[angle=-90,scale=0.35]{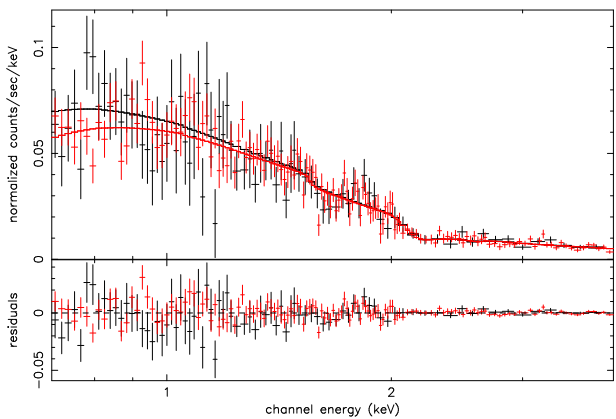}
\caption[source_spectra]{Summed spectra for all X-ray
sources, with the $2000$ data shown in black and the $2002$
data in red. The curves indicate the best-fit power-law model,
with index $\kappa=1.64$, obtained by fitting
simultaneously to both data sets. The lower panel shows
the residuals from this best-fit model.
\label{fig:source_spectra}
}
\end{figure}

The composite, background-subtracted source spectra for the $2000$ and
$2002$ datasets were fit simultaneously with XSPEC v11.2.0 (Arnaud 1996).
The spectral energy distribution is assumed to have a single photon 
power-law behaviour $\propto E^{-\kappa}$.
The Galactic hydrogen column density
is held fixed at $N_H=2.5\times 10^{20}$~cm$^{-2}$ in the fit; channels
with energy less then $0.7$ keV and greater than $4$ keV were excluded
from the fit. The spectra, best-fit model and residuals are shown in
Figure~\ref{fig:source_spectra}. The best-fit power-law exponent
is $\kappa = 1.64^{+0.047}_{-0.046}$ ($90\%$ confidence uncertainties), with
a reduced chi-squared of $\chi^2_{\nu}=1.032$ with $218$ degrees of freedom.
Irwin et~al. (2003) analyzed the composite spectra
of point sources within $3R_e$ for a sample of $15$ 
nearby galaxies, finding power-law exponents in the range
$1.45 \le \kappa \le 1.9$. Thus, our measured power-law exponent for
the LMXBs in M87 is typical of those found in other nearby galaxies.
This best-fit model is used to convert the observed counts to
unabsorbed luminosities,  $L_X$, over the range $0.3-10$ keV,
assuming that all the sources are at the distance of M87. The resulting
conversion factor is $1.4 \times 10^{41}$ erg count$^{-1}$.

\begin{figure}
\plottwo{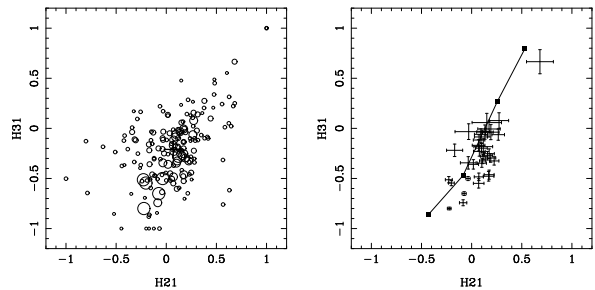}{f5b.eps}
\caption[hardness]{\textit{Left}: Hardness ratios, H21 and H31, for the full sample of
174 X-ray point sources. 
The size of the symbol for each source is proportional
to its luminosity in the range 0.3--10 keV. The mean hardness ratios are
$(\langle\mbox{H}21\rangle,\langle\mbox{H}31\rangle) = (0.07,-0.21)$,
as indicated by the cross.
\textit{Right}: Hardness ratios, H21 and H31, for all sources with 
$L_X > 1.5\times10^{38}$ erg s$^{-1}$. Error bars give $1\sigma$ uncertainties
in the ratios. The line shows predicted ratios for a power law with 
a hydrogen column density equal to 
$N_H=2.5\times10^{20}$ cm$^{-2}$ (Stark et~al. 1992); from top to bottom, the squares
correspond to the predictions for power law exponents
of 0, 1, 2 and 3.
\label{fig:hardness}
}
\end{figure}

We examined the crude spectral properties of the resolved sources by
calculating hardness ratios, which have the advantage of being 
measurable for even the faintest sources. Counts were calculated
for three distinct energy bands: a soft ($0.3-1$ keV) band denoted by
$S$, a medium band ($1-2$ keV) denoted by $M$, and a hard ($2-10$ keV)
band denoted by $H$. Following Sarazin et~al. (2000), the hardness
ratios, H$31$ and H$21$, are taken to be
\begin{equation}
\begin{array}{rcl}
H21 & \equiv & (M-S)/(M+S) \\
H31 & \equiv & (H-S)/(H+S) \\
\end{array}{}
\label{eq2}
\end{equation}
The distribution of hardness ratios is shown in Figure~\ref{fig:hardness}.
The sources occupy a diagonal swath in this plot, as is typical for LMXBs
(Sarazin et~al. 2000; Blanton et~al. 2002; Irwin et~al. 2003; Jeltema et~al.
2003; Sivakoff et~al. 2003). The mean location, at $(0.07,-0.21)$, is 
indicated by the cross.
It is apparent from this figure that the most luminous sources appear to have
the softest spectra, and a Wilcoxon test confirms this impression,
giving probabilities of 6\% and 0.6\%, respectively, that the  H$21$ and H$31$
values for sources with $L_X > 5\times10^{38}$~erg~s$^{-1}$ share
the same location. Although there are only six
objects with $L_X > 5\times10^{38}$~erg~s$^{-1}$, this result is 
consistent with that of Irwin et~al. (2003), who noted that
sources with $L_X > 10^{39}$~erg~s$^{-1}$ appear to be significantly softer.
As they remark, this trend might be a reflection of inverse dependence
between the emitted flux and spectral state exhibited by candidate black hole 
X-ray binaries in the Milky Way ({\it e.g.} Tanaka \& Lewin 1995;
Nowak 1995). Finally, we note that there
are two sources in the upper right corner of Figure~\ref{fig:hardness}
in which both H$21$ and H$31$ are equal to $1.0$; given the hardness of
these spectra, we suspect that these sources may be strongly absorbed AGN.

\section{Global Properties of the LMXB Population}

We now turn our attention to the observed properties of the LMXB population
as a whole ($i.e.$, spatial structure, luminosity function, and suitability
as a distance indicator). Our ultimate aim is to understand
the nature of the connection between LMXBs and GCs in M87, and to examine
the broader implications for LMXB formation. 

\subsection{Radial and Azimuthal Structure}
\label{sec:radprof}

\begin{figure}
\plotone{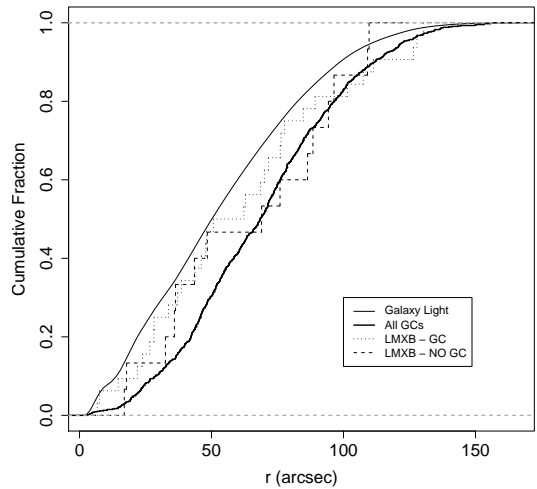}
\caption[radial_cum]{Comparison for the radial distribution of
galaxy light, globular clusters and LMXBs in M87. The heavy solid 
curve shows the cumulative distribution of 867 globular cluster
candidates in the ACS field that are brighter than $z_{850} = 22.8$.
The thin solid curve shows distribution of galaxy light within our
ACS field.  The dotted curve shows the cumulative distribution
of the 32 LMXBs in our ACS field that are
brighter than $L_X = 7\times10^{37}$~erg~s$^{-1}$ and coincide with
a globular cluster candidate. The dashed curve show the cumulative
distribution of the remaining sample of 15 LMXBs in 
the ACS field that are brighter than 
$L_X = 7\times10^{37}$~erg~s$^{-1}$ but have no GC counterpart.
\label{fig:radial_cum}
}
\end{figure}

How does the radial distribution of LMXBs in
M87 compare to that of its GCs and the underlying galaxy light?
In comparing the various profiles, we restrict ourselves to
X-ray point sources having $L_X > 7\times10^{37}$ ergs s$^{-1}$,
and to GC candidates that do not fall within the regions masked
during the X-ray point source detection procedure and 
having $z_{850} < 22.8$ mag (a total of 867 objects)
in order to guard against incompleteness effects. 
The heavy solid curve in Figure~\ref{fig:radial_cum}
shows the resulting cumulative distribution for the M87 GC system, calculated
directly from the GC catalog described in \S\ref{sec:optcat}. 

In principle, it should also be possible to measure the 
profile of the galaxy itself from our ACS images.
However, such an approach is undermined by the limited 
areal coverage of the ACS field and the fact that our
ACS images, which are centered on the galaxy's nucleus, provide limited
constraints on the background surface brightness. Given these problems,
we estimate the cumulative light distribution within the ACS field,
${\cal S}(r)$, by using the wide-field surface photometry of
Caon, Capaccioli \& Rampazzo (1990). For an annulus centered on the
galaxy, the fractional area falling within the ACS field (excluding
those regions masked during X-ray point source detection; see below) is
$f_a(r)$. The cumulative distribution of galaxy light is then
\begin{equation}
{\cal S}(r) =  2\pi r f_a(r)10^{-0.4\mu(r)} \\
\label{eq3}
\end{equation}
where $\mu(r)$ refers to the $B$-band surface brightness profile of
Caon et~al. (1990). Note that the implicit assumption of circular
symmetry in Equation~\ref{eq3} is quite reasonable for M87, which
has a luminosity-weighted mean ellipticity of
$\langle\epsilon\rangle \sim 0.05$ inside $r \sim 2\farcm75$ --- the
maximum galactocentric radius of our ACS field.

The thin solid curve in Figure~\ref{fig:radial_cum} shows
the cumulative profile of the galaxy light within the ACS field.
A KS test confirms the well known
result that the GC system of M87 has a shallower profile than 
the galaxy itself ($e.g.$, Grillmair et~al. 1986; Harris 1986). 
Also shown in Figure~\ref{fig:radial_cum} are
the cumulative distributions for two LMXB subsamples:
the dotted curve shows the distribution
for those sources that are associated with GC
candidates, while the dashed curve shows the distribution for
the remaining X-ray sources. In both cases, we plot only 
those X-ray sources that fall within the ACS field.
Since both samples are subject to the same selection effects,
it is straightforward to compare these distributions 
directly. A two sample KS test accepts the hypothesis that 
they were drawn from the same parent 
sample.

Comparing the X-ray point source samples to the galaxy
and GC profiles is more difficult.
Our {\it Chandra} image reveals the 
inner $\sim 40''$ of M87 to have a remarkably complex
structure in diffuse emission (Figure~\ref{fig:ximg}; see also
Figure~1 of Young, Wilson \& Mundell 2002 and 
Sparks et~al. 2004). Because of this complexity, it was
necessary to mask several problematic regions prior to object detection
(see \S\ref{sec:red}), limiting the region over which the 
various profiles can be compared. Perhaps as a consequence, a 
one-sample KS test accepts the hypothesis
that both X-ray samples (i.e., those point sources with, and without,
an associated GC candidate) were drawn from the 
same parent distribution as the galaxy light; moreover, a two-sample KS test
accepts the hypothesis that they were drawn from the same parent distribution
as the GC candidates. Stronger conclusions 
will require an expanded census of LMXBs but, given the brightness 
and complexity of the diffuse X-ray emission in the inner regions
of M87, the requisite observations will prove extremely challenging.

\begin{figure}
\plotone{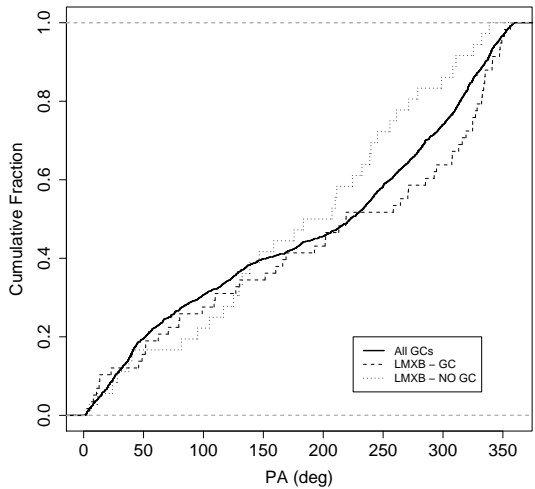}
\caption[ang_cum]{Comparison for the azimuthal distribution of
globular clusters and LMXBs in M87. The solid 
curve shows the cumulative distribution of all bona fide
GC candidates.
The dashed curve shows the cumulative distribution 
of the X-ray point sources that coincide with a GC
and the dotted curve shows the cumulative distribution
of the X-ray point dources that do not coincide with a GC.
\label{fig:az}
}
\end{figure}

We also explored the azimuthal distribution of the X-ray 
point source samples. In Figure~\ref{fig:az} we show
the cumulative distribution function for the X-ray point
sources that are associated with a GC and those that are not. We also
show the azimuthal distribution of the full GC sample 
(excluding regions that were masked in the X-ray point source
detection). It is clear from the figure
that the X-ray point sources associated with a GC
follow the distribution of the full GC sample, and this is
confirmed by a KS test. The X-ray point sources not associated
with a GC seem somewhat deviant, but a KS test accepts 
the hypothesis that the sample was drawn from the same distribution
as the full GC sample and the X-ray point sources
associated with a GC with p-values of $p_{KS}=0.19$
and $p_{KS}=0.1$ respectively. This exercise
reveals that the spatial distribution of 
X-ray point sources associated with a GC is representative 
of the full GC sample.

\subsection{Luminosity Function}
\label{sec:lx}

\begin{figure}
\plotone{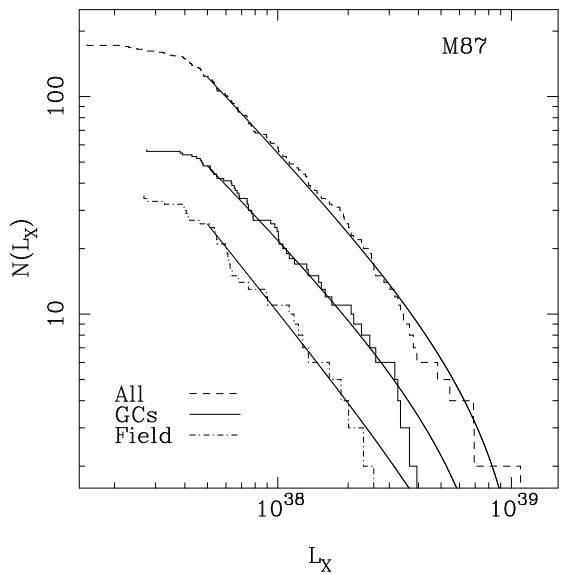}
\caption[lx_typ]{Observed $\log N(L_X)$, where
$N(L_X)$ is the number of X-ray sources with $L > L_X$.
The dashed line shows the distribution for all
sources within the S3 chip; the sources within the 
ACS field and associated with globular cluster are represented by the solid 
line; those sources within the ACS field but not associated
with a globular clusters are shown by the dot-dashed line. The smooth lines
are the best-fit power-law luminosity distributions of the form
$f \propto L_X^{-\gamma}$ with an upper cutoff such that $f=0$ for
$L_X > 10^{39}$~erg~s$^{-1}$. For the full sample of all sources, and
for the subsample of sources that not associated with a globular
cluster, the background contamination has been removed as in Giacconi
et~al. (2001).
\label{fig:lx_typ}
}
\end{figure}

The luminosity function of LMXBs is of considerable interest,
both as a rare constraint on the mass distribution of accreting
sources in external galaxies, and as a potential distance
indicator. Working with a sample of $\approx$ 80 LMXBs in NGC~4697, 
Sarazin et~al. (2001) showed that their cumulative luminosity
function was well described by a broken power-law, with a ``break"
at $L_b \approx 3.2 \times 10^{38}$~ergs~s$^{-1}$. Since
this is close to the Eddington luminosity for spherical
hydrogen accretion onto the surface of a $1.4 M_{\odot}$
neutron star ({\it e.g.} Shapiro \& Teukolsky 1983), 
Sarazin et~al. (2001) drew attention
to the possibility of using this feature as a standard
candle in distance estimation. Indeed, the use of a characteristic
luminosity in accreting neutron stars as a distance indicator 
dates to early studies of Galactic X-ray sources ($e.g.$,
Margon \& Ostriker 1973; van Paradijs 1978).

\begin{figure}
\plotone{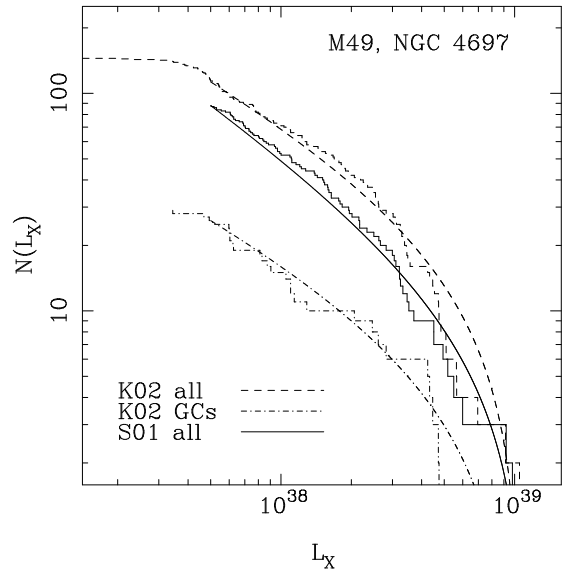}
\caption[lit_pow]{Same as Figure~\ref{fig:lx_typ}, except
for sources in M49 and NGC~4697. The dashed histogram shows 
the distribution of X-ray sources in M49 from the S3 chip
of Kundu et~al. (2002). The corresponding distribution for
sources in NGC~4697 is given by the solid histogram (Sarazin
et~al. 2001). The dot-dashed histogram shows the distribution
for sources in M49 that are associated with a globular cluster 
(Kundu et~al. 2002). The smooth curves are best-fit power-law
luminosity distributions of the form $f \propto L_X^{-\gamma}$
with an upper cutoff such that $f=0$ for $L_X > 10^{39}$~erg~s$^{-1}$. 
For the complete samples in both galaxies, the
background contamination has been removed as in
Giacconi et~al. (2001).
\label{fig:lit_pow}
}
\end{figure}

\subsubsection{Representation as a Broken Power-Law}

In Figure~\ref{fig:lx_typ}, we show the cumulative luminosity
function, $\log N(L_X)$, of X-ray sources in M87. Here,
$N(L_X)$ is the number of objects with luminosities in excess
of $L_X$. The dashed line shows this distribution for
the complete sample of sources --- a total of 174 objects. 
The two lower distributions show the distributions for those 
sources that fall within the ACS field: the solid line
indicates those sources that coincide with a GC, while the
dot-dashed line refers to sources with no associated GC.
Aside from normalization, the three distributions appear
remarkably similar; the abrupt flattening below
$L_X \sim 4\times10^{37}$~erg~s$^{-1}$ is probably due to incompleteness.
When fitting models to the luminosity functions, we have
corrected for background contamination using the 
background counts of Giacconi et~al. (2001) for the complete sample
and the subsample of objects that are not associated with GCs.

To facilitate comparison with previous work, we have fitted these
three distributions with broken power-laws of the form
\begin{equation}
\begin{array}{rcllll}
N(L_X) & \propto & L_X^{{\alpha}_1} & & {\rm for} & L_X < L_b \\
N(L_X) & \propto & L_X^{{\alpha}_2} & & {\rm for} & L_X \geq L_b \\
\end{array}{}
\label{eq4}
\end{equation}
To guard against incompleteness effects, we consider only those
source brighter than $5\times 10^{37}$~erg~s$^{-1}$.
The resulting values for ${\alpha}_1$, ${\alpha}_2$ and $L_B$ are
given in Table~\ref{tab:xlf}. 
Note that the break luminosities found here,
$L_B \sim (2-3)\times10^{38}$~erg~s$^{-1}$,
are similar to
those found in other early-type galaxies ($e.g.$, Sarazin et~al.
2001; Finoguenov \& Jones 2001; Kundu et~al. 2002). Although the
observed luminosity distribution is well described by this
particular choice of parameterization,
the data do not {\it require} a
broken power-law. As we now show, the data are equally well
represented by a single power-law distribution with an upper 
cutoff in luminosity.

\begin{figure}
\plotone{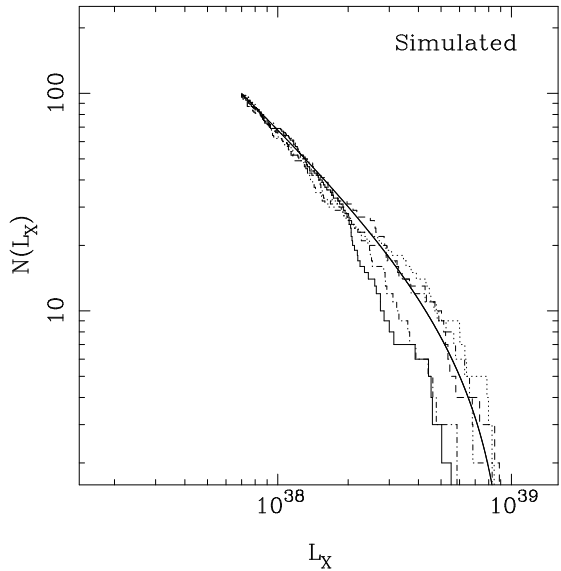}
\caption[eg]{The smooth solid line shows the expected behaviour of
$\log N(>L)$ for a sample of size $100$ drawn from a distribution of luminosities given
by $L^{-2}$ with an upper cut at $L=10^{39}$~erg~s$^{-1}$. The rest 
of the lines show the observed behaviour of $5$ random realizations
of size $100$ from that distribution. 
\label{fig:eg}
}
\end{figure}

\subsubsection{Representation as a Truncated Power-Law}

For each of the three samples of observed X-Ray luminosities,
whose corresponding cumulative distributions
are shown in Figure~\ref{fig:lx_typ},
we have fit single power-law distributions of the form
$L_X^{\gamma}$ for $L_{min}<L_X<L_{max}$, taking 
$L_{max}=10^{39}$~erg~s$^{-1}$ as the upper cutoff. As
before, we adopt $L_{min}=5\times 10^{37}$~erg~s$^{-1}$ to guard
against incompleteness at the faint end. The corresponding best-fit 
cumulative distributions are shown as the smooth curves. In each case,
a one-sample KS test reveals that the samples are consistent with
their being drawn from a single power law with $\gamma \sim -2.1$.
The best fit values for $\gamma$ are
given in Table~\ref{tab:xlf}\footnote{Note that the broken power law
is fitted using $N(L_X)$, whereas the truncated power law is fitted
using the observed samples of $L_X$, whose parent
distribution we denote by $f$. $N(L_X)$ is $(1-F)$ modulo
a normalization constant, where $F$ is the cumulative distribution of $f$. Thus,
as the bulk of the data lies at luminosities lower than the inferred
break luminosity $L_b$, we expect that $\alpha_1 \approx \gamma + 1$.
}. 

Is this true of the LMXB populations in other galaxies? In
Figure~\ref{fig:lit_pow}, we show the cumulative luminosity
functions of all LMXBs in NGC~4697 and M49 (using data from 
Sarazin et~al. 2001 and Kundu et~al. 2002, respectively).
Maximum likelihood fits to both datasets reveals that 
they are consistent with having been drawn from a single power-law
distribution with an upper cut at $L_X=10^{39}$~erg~s$^{-1}$. The
corresponding best-fit cumulative distributions, 
which have $\gamma = -1.63\pm0.14$ (NGC~4697), 
$\gamma = -1.58\pm0.12$ (M49, all X-ray point sources) and
$\gamma = -1.64\pm0.25$ (M49, X-ray sources associated with a GC), 
are given by smooth 
curves in Figure~\ref{fig:lit_pow}.
As this exercise demonstrates, it is dangerous to draw conclusions
about the underlying distribution, $f$, from the quantity $N(L_X)$,
particularly at the high-luminosity end, if $f$ is truncated
above some value.
Let us denote the cumulative distribution of $f$ by $F$, so that 
$\log N(L_X)$ is equivalent modulo a constant to $\log(1-F)$.
The slope $s$ of this function is given by $s=-f/(1-F)$.
If $f = 0$ for values of $L_X$ greater than
$L_{max}$ then, as we approach $L_{max}$, we generically 
expect $\lim_{L\rightarrow L_{max}} s = -\infty$, producing a dip
in the expected form of $\log(1-F)$. To give an example
germane to the present discussion, let us assume that
$f=L_X^{-2}$ with $f=0$ for $L_X>10^{39}$~erg~s$^{-1}$. 
The corresponding distribution $N(L_X)$ is
shown as the smooth curve in Figure~\ref{fig:eg}. For comparison,
the histograms in this figure show five simulated
datasets, each consisting of one hundred objects. Because the
parent distribution is cut beyond a maximum $L_X$ some simulated
datasets have an apparent break, even though the parent distribution has
no characteristic scale to distinguish the two regimes.

Based on the evidence presented above, we conclude that there
is no compelling evidence for two fundamentally different accretor
populations in M87, M49 or NGC~4697, and that a single power-law distribution
(truncated  above $\sim 10^{39}$~erg~s$^{-1}$) provides an adequate
description of the LMXB populations in all three galaxies.
Moreover, we have shown that
the apparent ``breaks" at $L_X \sim (1-4)\times10^{38}$~erg~s$^{-1}$
may be an artifact of this distribution. Our conclusions are in
agreement with those of Sivakoff et~al. (2003), who 
found that the luminosity distribution of LMXBs in NGC~4365 and
NGC~4382 could be better modeled by a power law having an upper
cutoff at $L_X\sim10^{39}$~erg~s$^{-1}$. This also appears to be
the case in the Milky Way: Grimm, Gilfanov \& Sunyaev (2001) find
a truncated power law to be an excellent representation of the
Galactic LMXB luminosity distribution.

In retrospect, the lack of a characteristic luminosity scale should
perhaps come as no surprise since the Eddington luminosity, $L_E$,
is usually computed under very particular assumptions: namely, 
spherical accretion of pure ionized hydrogen and Thomson scattering.
This is clearly an idealized situation, and there are 
various ways in which an accreting neutron star
can exceed this naive estimate: $e.g.$, unusual chemical
abundance, formation of a super-critical disk, radiation in the form
of relativistic jets and the presence of strong magnetic fields 
(see, $e.g.$, the discussion in Grimm et~al. 2001 and references
therein). 
Still another effect that would blur a characteristic scale 
is that the observed values of $L_X$ will be affected
by disk obscuration and scattering.
Irwin et~al. (2003) find essentially
no sources with $L_X > 2\times 10^{39}$~erg~s$^{-1}$ in their
study of the LMXB populations of $15$ early-type galaxies, and our
results are in line with theirs.
Thus, it is likely that the luminosity function of LMXBs
in most early-type galaxes is truncated at
$L_{max} \sim (1-2)\times10^{39}$~erg~s$^{-1}$.
This upper cutoff is probably not universal.
Indeed,
a significant number of very luminous sources 
($L_X\lae10^{40}$~erg~s$^{-1}$) have been observed in some
early-type galaxies, such as NGC~720 (Jeltema et~al. 2003).
In NGC~720, the spatial distribution suggests that the
more luminous sources could arise from a younger stellar
population whose formation was perhaps triggered by a recent merger.

\begin{figure}
\plotone{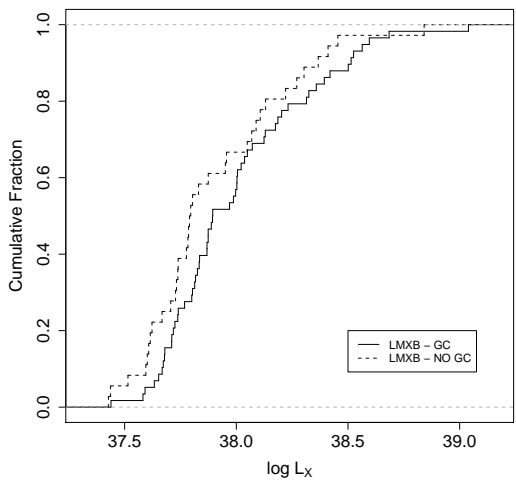}
\caption[lx_cum]{Normalized, cumulative luminosity functions
for LMXBs within the ACS field. The solid curve shows the distribution
of LMXBs that are associated with a GC; LMXBs that are not associated
with GCs are indicated by the dashed curve.
\label{fig:lx_cum}
}
\end{figure}

\subsubsection{Environmental Dependence}

Do the LMXBs in GCs share the same luminosity function as those
which are not associated with GCs?
There have been conflicting claims about such an
environmental dependence in the LMXB luminosity function. In their
study of NGC~1399, Angelini et~al. (2001) found the LMXBs in GCs
to be, on average, more luminous than those which are not associated with GCs.
On the other hand, Kundu et~al. (2002) found the two populations in
M49 to have indistinguishable luminosity functions, and  Sarazin et~al. (2003)
reached a similar conclusion based on their analysis of LMXBs 
in four early-type galaxies.

In Figure~\ref{fig:lx_cum} we show the normalized, cumulative luminosity 
function for LMXBs within the ACS field. The solid
curve shows the distribution of LMXBs that coincide with a GC, while the
dashed curve shows those LMXBs that do not. The LMXBs associated with
GCs are  slightly brighter on average, with a mean luminosity of
$\langle L_X \rangle=(1.4\pm0.2)\times10^{38}$~erg~s$^{-1}$. The mean
luminosity of LMXBs that are not associated with GCs is
$\langle L_X \rangle=(1.1\pm0.2)\times10^{38}$~erg~s$^{-1}$.
The difference, however, is not significant at greater
than $99\%$ confidence, as 
KS and Wilcoxon sum rank test accept the hypothesis
that the data were drawn
from the same distribution with p-values of
$p_{KS}=0.1$ and $p_{wilcox}=0.1$, respectively. Thus, at least
in the case of M87, we find no support for the claim that the LMXBs
that are associated with GCs are significantly brighter than
those which are not. These findings are consistent with the suggestion
that essentially all LMXBs in early-type galaxies may have first
formed in GCs (White, Sarazin \& Kulkarni 2002) and were subsequently
ejected or dispersed into the general field.

\section{The Relation Between Low Mass X-Ray Binaries and Globular Clusters}
\label{sec:plmxb}

\subsection{The Efficiency of LMXB Formation}
\label{sec:eff}

The most basic characterization of the probability 
that a GC harbors an LMXB is the
overall probability, $p_X$,  that a GC will host a LMXB candidate.
Restricting ourselves to those sources in the ACS field
of view, we find $p_X=0.036\pm0.005$\footnote{Because of incompleteness, this number is
a slight overestimate, as we are missing faint
GCs that will not contribute many LMXBs (see \S~\ref{sec:lumin}). 
We can estimate the magnitude of
the bias by assuming that the GC luminosity function 
$\Phi(m)$ is represented by a Gaussian with $\sigma=1.4$
and turnover magnitude $M_V=-7.4$ (Harris 2001). Using this form,
the luminosity function $\phi$ of GCs associated with an LMXB will satisfy
$\phi(m) \sim L^{0.89}\Phi(m)$ (\S~\ref{sec:lumin}), where
$L$ is the GC luminosity. Normalizing the distributions
by the number of GCs brighter than the turnover, we find that
after
correcting for incompleteness, $p_X\sim0.034$. This is
very similar to the directly observed value of
$p_X=0.036\pm0.05$, so we conclude that any bias is very small.
}.
This overall probability has been found
to be in the range $2-4\%$ in a wide variety
of galaxy types (see, e.g., Sarazin et~al. 2003 and references therein), 
providing a very uniform characteristic
of the connection between GCs and LMXBs in GCs. 
Another basic quantity is the fraction of X-ray point sources $f_{\rm X-GC}$
that reside in a GC; for M87 this is $f_{\rm X-GC}=0.62$. Observed values
of $f_{\rm X-GC}$ vary substantially from galaxy to galaxy, from $\sim0.2$  for 
sources in the central region of M31 
(Primini, Forman \& Jones 1993) to $\sim 0.7$ for NGC~1399 
(Angelini et~al. 2001),
and it has been proposed that the available observations indicate a systematic 
increases along the Hubble sequence from late to early types
(Sarazin et~al. 2003).
We now turn to consider the dependence of $p_X$ on various factors.

\subsection{Dependence on Metallicity}
\label{sec:metal}

In the Galaxy and M31, there is a tendency for LMXBs in GCs to
appear preferentially in metal-rich GCs (Grindlay 1993; 
Bellazzini et~al. 1995), but the limited samples sizes
($i.e.$, just 13 and 19 LMXBs, respectively, with
$L_X > 10^{36}$~erg~s$^{-1}$; Verbunt 2001, White \& Angelini 2001)
have precluded definite conclusions. However, new observations
have confirmed the trend with larger
samples of LMXBs in NGC~1399 (Angelini et~al. 2001), 
NGC~4472 (Kundu et~al. 2002) and NGC~5128 (Minniti et~al. 2004). 
Kundu et~al. (2002) find that metal-rich GCs are, on average, 
$\sim 3$ times as likely to harbor LMXBs than their 
metal-poor counterparts.

\begin{figure}
\plotone{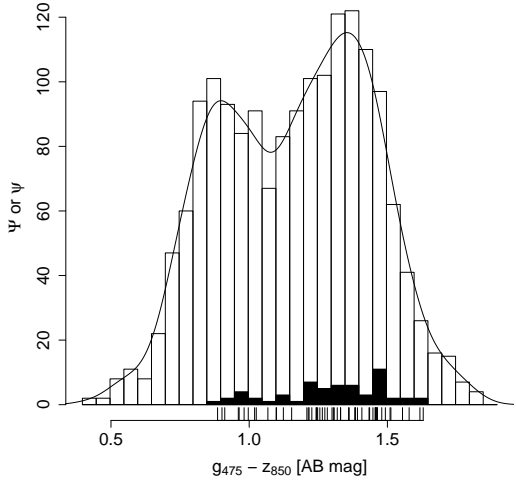}
\caption[col_dist]{Color histograms for the full sample 
of globular clusters (open bars) and for the sample
of globular clusters
which coincide with an X-ray point source (filled bars).
The curve is a normal kernel density estimate of the color
distribution, which we denote by $\hat{\Psi}(g_{475}-z_{850})$.
The tickmarks show the colors of individual GCs
associated with X-ray point sources.
\label{fig:col_dist}
}
\end{figure}

In Figure~\ref{fig:col_dist}, we show the ($g_{475}-z_{850}$) color distribution
for the full sample of 1688 GCs in our ACS field, along with the
corresponding distribution for the 58 GCs which coincide with an
X-ray point source.
It is apparent that LMXBs show a marked preference
for metal-rich GCs, and this is confirmed with two different statistical
tests: a two sample KS test rejects the hypothesis
that the distributions were drawn from the same distribution
with a p-value of $p_{KS}=3\times 10^{-4}$, while
a Wilcoxon rank sum test rejects the hypothesis
that the parent distributions have the same location
with a p-value of $p_{wilcox}=2\times 10^{-4}$.

To divide the color distribution into metal-rich and
metal-poor subsamples, we use the KMM algorithm (Ashman, Bird \& Zepf 1994).
This algorithm performs the division based on an \textit{a posteriori}
likelihood using two Gaussians to model the color distribution.
KMM gives a dividing color of ($g_{475}-z_{850}$) = 1.121, which we adopt as the
division between the metal-rich and metal-poor subpopulations in M87.
With the subpopulations defined in this way, we find a probability of
$p_{X,MR} = 5.1\pm0.7$\% that a given metal-rich cluster will contain
a LMXB; this should be compared to the value of
$p_{X,MP}=1.7\pm0.5$\% found for the metal-poor GCs.\footnote{In both case,
the uncertainties are computed assuming a Bernoulli distribution.}
We conclude that metal-rich GCs are $3\pm 1$ times more
likely to contain LMXBs than metal-poor GCs.

\begin{figure}
\plotone{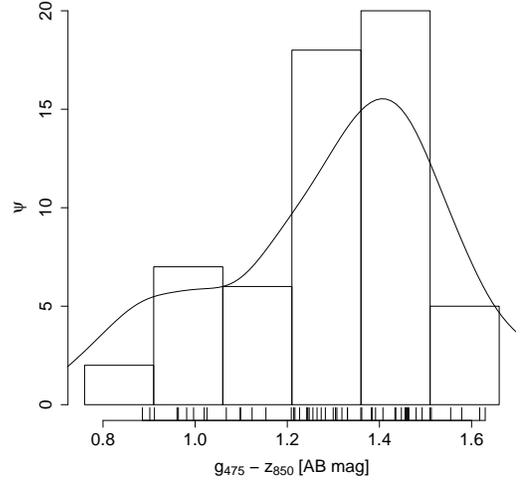}
\caption[col_beta]{Histogram of $(g_{475}-z_{850})$ colors for 58 globular cluster
in M87 which coincide with an X-ray point source. The curve shows a
model color distribution of the form $\psi \sim 10^{\beta(g_{475}-z_{850})}\hat{\Psi}$,
with $\beta=0.77$ (see text for details).
The tickmarks show the colors of individual objects.
\label{fig:col_beta}
}
\end{figure}

\begin{figure}
\plotone{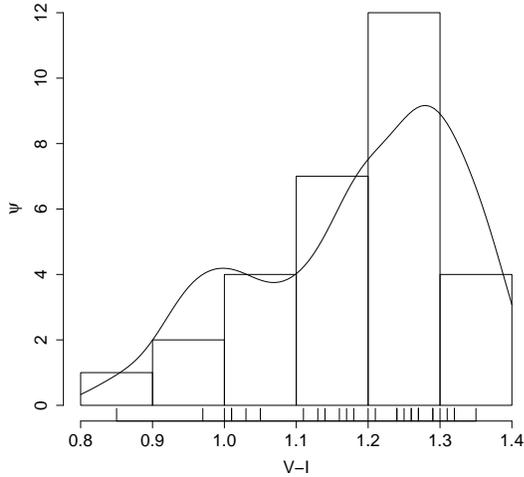}
\caption[mac_col_beta]{Histogram of $(V-I)$ colors
for globular clusters in M49 that coincide with an X-ray point
source, according to Maccarone et~al. (2002). The curve
shows a model color distribution of the
form $\psi_{VI} \sim 10^{\beta_{VI}(V-I)}\hat{\Psi}_{VI}$, with
$\beta_{VI}=1.2$ (see text for details).
The tickmarks show the colors of individual objects.
\label{fig:mac_col_beta}
}
\end{figure}

To get a quantitative estimate of the dependence of $p_X$ on metallicity,
we model the color distribution of GCs which contain LMXBs as
\begin{equation}
\begin{array}{rcl}
\psi & \sim & 10^{\beta(g_{475}-z_{850})} \Psi, \\
\end{array}{}
\label{eq5}
\end{equation}
where $\Psi$ is the color distribution for the full sample of GCs. 
To estimate $\Psi$, we use a kernel density estimate 
(Silverman 1986) which we denote by $\hat{\Psi}$, and
determine $\beta$ via a maximum likelihood fit
of the function 
\begin{equation}
\begin{array}{rcl}
\psi & \sim & 10^{\beta(g_{475}-z_{850})}\hat{\Psi} \\
\end{array}{}
\label{eq6}
\end{equation}
to the color distribution of the subsample of GCs that contain LMXBs.
The result is $\beta=0.77\pm 0.22$, and the 
distribution predicted by equation~(6) is compared to the observed one 
in Figure~\ref{fig:col_beta}. To find the dependence
of $p_{X}$ on metallicity, it would be best to use an empirical 
determination of the relation between 
$(g_{475}-z_{850})$ and [Fe/H], but unfortunately no such relation is
available in the literature to the best of our knowledge.
As a substitute, we use the models of
Bruzual \& Charlot (2003) to find the relation
between $(g_{475}-z_{850})$ and [Fe/H]. Using the data
listed in Table~\ref{tab:bc}, we find a best-fit linear relation of
$(g_{475}-z_{850})\sim (0.38\pm0.05)\mbox{[Fe/H]}+(1.62\pm0.06)$, with an rms
scatter of roughly 0.1 mag\footnote{ The relation between
[Fe/H] and $(g_{475}-z_{850})$
is slightly better described by a quadratic relation,
$(g_{475}-z_{850})\sim (0.124\pm0.017)\mbox{[Fe/H]}^{2}
+(0.622\pm0.034)\mbox{[Fe/H]}+(1.620\pm0.015)$. If we use
this relation, we would find that
$\psi \sim 10^{0.0992\mbox{{\scriptsize [Fe/H]}}^{2} + 0.4976\mbox{{\scriptsize [Fe/H]}}}\hat{\p\Psi}$.
We prefer to use the linear relation because it adequately represents
the relation between [Fe/H] and $(g_{475}-z_{850})$ in the range
where the GCs used to derive $\psi$ lie 
($0.9 \lae (g_{475}-z_{850}) \lae 1.6$) and because it allows
us to cast $\psi$ simply in terms of a power law in $Z$.
Although we do not attach any special physical 
significance to a power law form, it describes
the observed  trend with relative simplicity.
}. Using this relation, we find
\begin{equation}
\begin{array}{rcl}
\psi & \sim & 10^{(0.32\pm0.10)\mbox{{\scriptsize [Fe/H]}}}\hat{\Psi}\\
     & \sim & (Z/Z_{\odot})^{(0.32\pm0.10)}\hat{\Psi} \\
\end{array}{}
\label{eq7}
\end{equation}
We also reanalyzed the metallicity dependence of LMXBs in the GC
system of M49, using the ($V-I$) colors presented in Maccarone
et~al. (2002). Fitting a model of the form of equation~(6), we
find $\beta_{VI}=1.2\pm0.5$. The empirical color-metallicity
relation of Barmby et~al. (2000) then gives
\begin{equation}
\begin{array}{rcl}
\psi_{VI} & \sim & 10^{(0.19\pm0.08)\mbox{{\scriptsize [Fe/H]}}}\hat{\Psi}_{VI}\\
          & \sim & (Z/Z_{\odot})^{(0.19\pm0.08)}\hat{\Psi}_{VI} \\
\end{array}{}
\label{eq8}
\end{equation}
consistent with our findings for M87. The color distribution for
the M49 data, along with the best-fit model, is shown in
Figure~\ref{fig:mac_col_beta}.

\subsection{Dependence on Luminosity}
\label{sec:lumin}

In addition to metallicity, luminosity plays an important role
in determining whether a given GC will contain a LMXB (in the
sense that LMXBs reside preferentially in the most luminous
clusters).  This has been observed to be the case in NGC~1399
(Angelini et~al. 2001), M49 (Kundu et~al. 2002), four
early-type galaxies analyzed in Sarazin et~al. (2003) and NGC~5128 
(Minniti et~al. 2004). Given
that our ACS observations of M87 define the deepest, most
complete sample of GCs yet assembled for any galaxy, we now
examine the dependence of $p_X$ on luminosity in M87.

\begin{figure}
\plotone{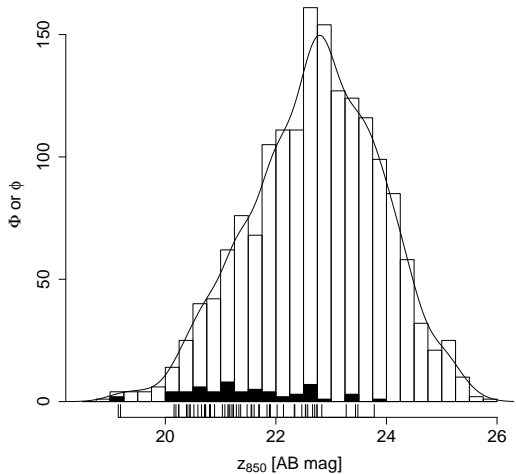}
\caption[lf]{$z_{850}$-band luminosity function for the
full sample globular cluster candidates (empty bars) and for those candidates
that coincide with an X-ray point source (filled bars). The curve
is a normal kernel density estimate of the luminosity
function, which we denote by $\hat{\Phi}(m)$.
The tickmarks show the magnitudes of individual GCs
associated with X-ray point sources.
\label{fig:lf}
}
\end{figure}

In Figure~\ref{fig:lf}, we show the $z_{850}$-band
luminosity function for the full sample of 1688 GCs within
the ACS field of view (open histogram), along with the corresponding
luminosity function for those 58 GCs which coincide with an X-ray
point source (filled histogram). It is clear that, in agreement
with previous findings, the LMXBs are associated preferentially
with the brighter GCs. A two sample KS test rejects the hypothesis
that the two datasets were drawn from the same parent distribution
and a Wilcoxon rank sum test rejects the hypothesis
that the parent distributions share the same location; the
respective p-values are $p_{KS}=7\times 10^{-11}$ and 
$p_{wilcox}=7\times 10^{-14}$.

Both Kundu et~al. (2002) and Sarazin et~al. (2003)
argue that the data are consistent with the probability
per unit luminosity being constant. 
To represent the luminosity function of those
GCs that coincide with LMXBs, $\phi(m)$, we will adopt a
probability density of the form
\begin{equation}
\begin{array}{rcl}
\phi(m) & \sim & L^{\alpha}\Phi (m) \\
        & \sim & 10^{-0.4\alpha m} \Phi(m) \\
\end{array}{}
\label{eq9}
\end{equation}
where $\Phi(m)$ is the distribution of GC magnitudes.
At this point, we could model $\Phi(m)$ parametrically
(e.g., by using a Gaussian and taking care to model
the effects of incompleteness). However, we prefer to
approximate $\Phi(m)$ using a normal kernel density estimate
(Silverman 1986) since this approach does not require us to
make any assumptions about the true parent distribution, while
at the same time, the effects of incompleteness are taken
into account. 

\begin{figure}
\plotone{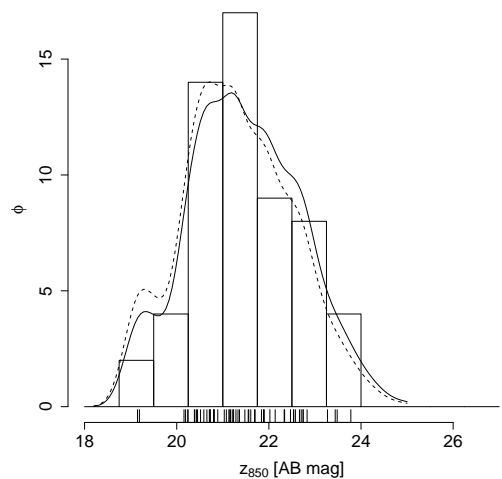}
\caption[xlum]{$z_{850}$ band luminosity function for globular cluster
candidates with an associated X-ray point source.
The solid and dashed curves are model distributions of the
form $\psi \sim L^\alpha \hat{\phi}$, with
$\alpha=0.89$ and $\alpha=1$, respectively. 
The tickmarks show the magnitudes of individual objects.
\label{fig:xlum}
}
\end{figure}

Denoting the kernel density estimate by $\hat{\Phi}(m)$, we 
determine the parameter $\alpha$ via a maximum likelihood fit of the 
function $10^{-0.4\alpha m} \hat{\Phi} (m)$ to the
observed distribution, $\phi(m)$.  The result is $\alpha=0.89\pm0.12$,
which is consistent with the 
results of Sarazin et~al. (2003), who found
that the luminosity function of GCs with associated LMXBs
was consistent with the form
$\phi(m) \sim  L \Phi(m)$. In Figure~\ref{fig:xlum},
we show the luminosity function of GCs with an associated LMXB, along 
with the best-fit LMXB luminosity functions $\phi \sim L^{0.89} \hat{\Phi}(m)$
(solid curve) and $\phi \sim L\hat{\Phi}(m)$ (dashed curve).

\subsection{Dependence on Encounter Rates}
\label{sec:enc}

For the first time for a galaxy outside the Local Group, we are able to 
examine possible variations of $p_X$ with GC structural parameters. The
number of LMXBs 
is expected to depend on these 
parameters since the binaries responsible for the X-ray emission
are likely to have a dynamical 
origin. That is to say, compact binaries in which one of the
components is a neutron star are probably {\it not} primordial in
nature, but have likely formed as a result of the tidal capture 
of neutron star, or by exchange interactions with pre-existing
binaries (Verbunt 2002).

The encounter rates, $\Gamma$, for both tidal 
capture and exchange interactions satisfy

\begin{equation}
\Gamma \propto \frac{\rho_0^2r_c^3}{v} 
\label{eq10}
\end{equation}

\noindent where $\rho_0$ is the central mass density of the GC, 
$r_c$ is the core radius and $v$ is the relative velocity of the encounter
(Hut \& Verbunt 1983).
An estimate of $v$ can be
obtained from the velocity dispersion of the cluster, 
which in turn is related to the central density and core radius
via the virial theorem, $\sigma \propto r_c \sqrt{\rho_0}$. 
To be sure, other factors will affect the
encounter rates for a given cluster, such as 
the rate at which binaries formed by these 
mechanisms are subsequently disrupted, the mass function in the cluster core, 
the binary fraction, and the period 
distribution of the binaries (Verbunt 2002). In this section, however,
we concentrate on the explicit dependence of $\Gamma$
on structural parameters; the role of other factors, which may 
themselves depend on metallicity 
and structural parameters, will be examined 
in \S\ref{sec:implications}.
In the absence of more detailed information, we will
first test the simple scenario in which $p_X$
is proportional to

\begin{equation}
\Gamma \equiv \rho_0^{1.5}r_c^2.
\label{eq11}
\end{equation}

To compute $\Gamma$,
we use the relations between core radius and half-light radius,
$\log {\cal R}(c) \equiv \log (r_h/r_c)$, and between
the dimensionless luminosity,
$\log {\cal L}(c) \equiv \log (L/j_0r_c^3)$, and
King concentration parameter, 
$c \equiv \log (r_t/r_c)$, which are given in Appendix~B 
of McLaughlin (2000). Here, $j_0$ is the central
luminosity density, $r_h$ is the half-light radius, 
$r_c$ is the core radius,\footnote{Note that 
in the notation of McLaughlin (2000) the core radius is referred
to as $r_0$.} $r_t$ is the tidal radius and $L$
is the cluster's $V$-band luminosity.
Explicitly, if $m$ is the $z_{850}$-band magnitude, $DM = 31.03$ the distance
modulus of M87 (Tonry et~al. 2001), $M_{V,\odot}= 4.84$ 
the absolute V-band magnitude of the Sun 
(Vandenberg \& Bell 1985), $j_{0}$ 
the central $V$-band luminosity density, and $\Upsilon_V$
the $V$-band mass-to-light ratio, then the central
mass density, in units of $M_{\odot}$~pc$^{-3}$, 
is given by the expression

\begin{equation}
\rho_0 = \Upsilon_V j_{0} 
= \Upsilon_V \frac{10^{-0.4(m - DM - M_{V,\odot} 
+ c_V)}}{(r_h/{\cal{R}})^3{\cal{L}}}
\label{eq12}
\end{equation}

\noindent where the core radius is given by
$r_c=r_h/{\cal{R}}$ and $c_V$ is the color term needed to 
convert from $z_{850}$-band to $V$-band luminosity. 
The $V$-band mass-to-light ratios of Galactic GCs are consistent with
a constant value of $\Upsilon_V=1.45$ in solar units (McLaughlin 2000),
 which we  henceforth adopt for the M87 GCs. In general, the color term $c_V$ is 
ill-defined as it will depend on age and metallicity. We  
assume that the bulk of the GCs in M87 are old and coeval,
as is suggested by spectroscopic and 
photometric age measurements 
(Cohen, Blakeslee \& Ryzhov 1998; Jord\'an et~al. 2002;
Kissler-Patig et~al. 2002). Assuming a mean age of $13$ Gyr
(Cohen et~al. 1998), we obtain the relation between ($g_{475}-z_{850}$) 
and ($V-z_{850}$) using the population synthesis models of 
Bruzual \& Charlot (2003). For each ($g_{475}-z_{850}$) color,
we linearly interpolate (or extrapolate, if necessary) to find
the corresponding  ($V-z_{850}$) at  each of the metallicities in the Bruzual \& 
Charlot (2003) models. The data used to define the color-color relation 
are listed in Table~\ref{tab:bc}.

\begin{figure}
\plotone{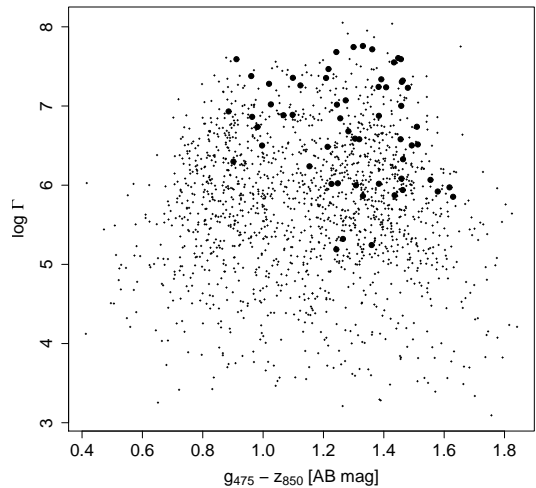}
\caption[col_gamma]{Distribution of $(g_{475}-z_{850})$ color as a function 
of encounter rate, $\Gamma \equiv \rho_0^{1.5}r_c^2$,
for the full sample of 1688 globular clusters (small crosses). Large circles
indicate the 58 globular clusters which contain a LMXB.
\label{fig:col_gamma}
}
\end{figure}

For the full sample of 1688 GCs, we find a mean encounter rate of
$\langle \log \Gamma\rangle = 5.8\pm 0.02$. For comparison, the mean
encounter rate for the subsample of 58 GCs which coincide with
a LMXB is $\langle \log \Gamma\rangle = 6.73\pm 0.09$.
A two sample KS test 
rejects the hypothesis that the distribution of $\Gamma$ parameters
for these two samples arise from the same parent 
distribution and a Wilcoxon rank sum test 
rejects the hypothesis that they have the same 
location; the respective p-values are $p_{KS}=3\times 10^{-11}$ and 
$p_{wilcox}=3\times 5^{-15}$.
{\it This constitutes the strongest evidence to date that encounter rates 
play a key role in determining } $p_X$. 

This finding is also consistent
with the results of Pooley et~al. (2003) and Heinke et~al. (2003).
Pooley et~al. (2003) find that $\Gamma$
is the main factor in determining the number of
close X-ray binaries in Galactic GCs. Specifically, they find
$p_X\propto \Gamma$ after restricting their analysis to the subset of GCs
which contain bonafide LMXBs
(see also Verbunt \& Hut 1987), although 
their conclusions are hampered by the limited sample size.
Heinke et~al. (2003) find that the population of  
{\it quiescent} LMXBs in Galactic GCs is consistent with their dynamical
origin as indicated by $\Gamma$.
In M31, Bellazzini et~al. (1995) have shown that the central
density of GCs which host LMXBs is higher than the mean
central density of M31 GCs, which also points to the importance
of encounter rates in determining the presence of LMXBs in 
GCs.\footnote{Note that the variations of
$\Gamma$ are driven mainly by $\rho_0$ rather than $r_c$ -- 
the former quantity varies by $\sim 3$ orders of magnitude and has a strong
dependence on luminosity, while the latter quantity varies by $\sim 1$ order of magnitude
and has a modest dependence on color or luminosity.}
Taken together, there seems to be little doubt that
exchange interactions and tidal captures in GCs are largely responsible
for the production of LMXBs. 

In Figure~\ref{fig:col_gamma}, we plot the measured
values of $\Gamma$ against ($g_{475}-z_{850}$) color for the full sample
of GCs (small symbols), along with the subsample of GCs which contain
LMXBs (large symbols).
There is a clear tendency for the latter GCs to have
higher than average encounter rates, and it is apparent 
that no obvious correlation exists between $\Gamma$ and
GC color (recall from \S~\ref{sec:metal} that
metallicity is an important factor in determining
$p_X$). Since the encounter rate and metallicity are uncorrelated,
we can assume they are independent so that 
$p_X \propto p_1(\Gamma) p_2(\mbox{[Fe/H]})$.

\begin{figure}
\plotone{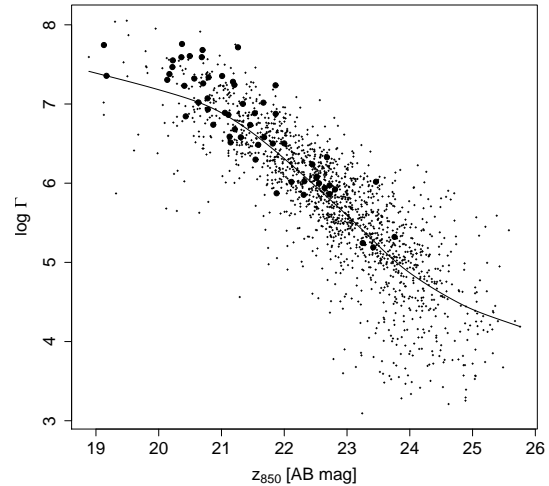}
\caption[mag_gamma]{Plot of $z_{850}$-band magnitude versus $\log \Gamma$
for all globular clusters (small crosses). Large circles
indicate those clusters that coincide with a LMXB.
The curve shows a smoothing spline fit to the full dataset.
\label{fig:mag_gamma}
}
\end{figure}

In \S\ref{sec:lumin}, we showed that there is a strong correlation between
$p_X$ and GC luminosity. In Figure~\ref{fig:mag_gamma}, we
plot $\Gamma$ versus $z_{850}$-band magnitude for the full sample
pf GCs (small symbols), as well as for the those GCs which are
associated with a LMXB (large symbols). There is a clear
tendency for $\Gamma$ to increase with increasing luminosity, which
raises the possibility that the trend discussed in \S\ref{sec:lumin} is 
a consequence of a more fundamental correlation between $\Gamma$ and 
luminosity. In fact, we can {\it predict} the luminosity distribution of GCs which
contain LMXBs, $\phi (m)$,  under the assumption
that $p_X \propto \Gamma$. Since there is no correlation between
luminosity and color in the M87 GC system (Whitmore et~al. 1995;
Harris, Harris \& McLaughlin 1998), we can safely ignore the 
metallicity dependence of $p_X$ when making this prediction.
To find the behaviour
of $\Gamma$ as a function of $z_{850}$-band magnitude, $m$, we fit a cubic 
smoothing spline to the data in Figure~\ref{fig:mag_gamma}.\footnote{A 
smoothing spline minimizes over all functions $f$ with continous second 
derivatives a compromise between the fit and the smoothness of the form
$\sum (y_i-f(x_i))^2+\lambda\int(f''(x))^2\,dx$ 
where $\{x_i,y_i\}$ are the data and $\lambda$ controls
the degree of smoothness and is chosen via cross-validation
(Hastie \& Tibshirani 1990; Green \& Silverman 1994).}
The resulting relation, $\Gamma(m)$,
is shown in Figure~\ref{fig:mag_gamma} as the smooth curve.
Using a kernel density estimate, $\hat{\Phi} (m)$, of the magnitude distribution
for the full sample of GCs, the predicted distribution for the subsample of GCs that contain
LMXBs should then satisfy $\phi (m) \sim \Gamma (m) \hat{\Phi} (m)$.

\begin{figure}
\plotone{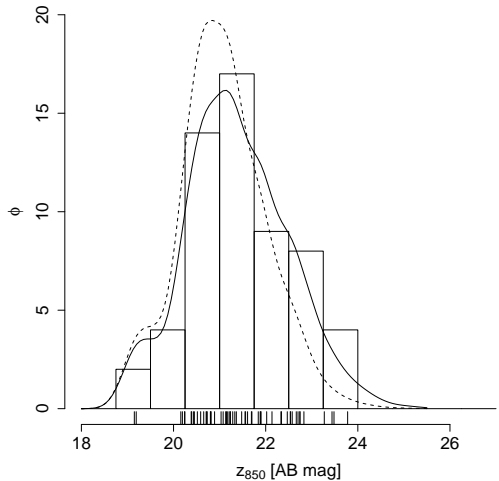}
\caption[pred_gamma_mag]{$z_{850}$-band luminosity function for globular
clusters with an associated  LMXB. The solid
curve is a model distribution of the form $\phi \sim \Gamma_{\alpha=1.08} \hat{\Phi}$,
while the dashed curve shows a model of the form $\phi \sim \Gamma \hat{\Phi}$.
The tickmarks show the magnitudes of individual objects.
\label{fig:pred_gamma_mag}
}
\end{figure}

This predicted distribution is shown as the 
dotted line in Figure~\ref{fig:pred_gamma_mag}. 
We stress that no fitting has been done in making this comparison; the only
assumption is that $p_X \propto \Gamma$. The predicted
distribution is in reasonable agreement with the observed distribution, although
it seems to underpredict somewhat the  number of faint GCs. A one-sample KS 
test gives $p_{KS}=0.07$, which does not allow us to reject the hypothesis 
that the observed distribution is explained by the assumption 
$p_X \propto \Gamma$ with better than $99\%$ confidence.

Rather than {\it assume} $p_X \propto \Gamma$, we now consider  
the dependence of $p_X$ on a more general quantity, $\Gamma_{\alpha}$,
which satisfies

\begin{equation}
\Gamma_{\alpha} \propto \rho_0^{\alpha}r_c^2
\label{eq13}
\end{equation}

\noindent where $\alpha$ is a free parameter that is determined
by fitting to the observed magnitude distribution of those GCs which
contain LMXBs. Such a form for
the encounter rate has been considered previously by 
Johnston, Kulkarni \& Phinney (1992)\footnote{Johnston et~al. (1992) 
and Johnston \& Verbunt (1996) assume
$\Gamma_\alpha \propto \rho_0^{\alpha_J} M_c 
\propto \rho_0^{1+\alpha_J}r_c^3$. Using the rough
correlation $r_c\propto \rho_0^{-0.2}$ (McLaughlin 2000) it follows that
$\alpha \sim \alpha_J + 0.8$ relates our $\alpha$ to theirs.},
who find $\alpha \sim 1.3$ from an analysis of pulsars in Galactic GCs, 
and by Johnston \& Verbunt (1996) who also
find $\alpha\sim 1.3$ from an analysis of low-luminosity
X-ray sources in Galactic GCs.
By adding this additional parameter, we are able to account for possible
variations in other factors which may
influence the probability of forming LMXBs, such as  
variations in the initial mass function (IMF) or systematic variations
in the relative importance of tidal captures
and binary-neutron star exchanges ({\it cf.} Grindlay 1996) . 
We would like $\Gamma_\alpha$ to share with $\Gamma$ the property
of being uncorrelated with $(g_{475}-z_{850})$ so as to be able to
consider them independent variables and thus 
to write $p_X=p_1(\Gamma_\alpha)p_2(\mbox{[Fe/H]})$. 
We therefore define $\Gamma_\alpha$ as 
\begin{equation}
\Gamma_\alpha \equiv  \rho_0^{\alpha}r_c^2 10^{(0.46-\alpha 0.33)(g_{475}-z_{850})},
\label{eq14}
\end{equation}
where we have made use of the fact that  $r_c^2 \sim 10^{(-0.46\pm0.06)(g_{475}-z_{850})}$ 
and $\rho_0 \sim 10^{(0.33\pm0.09)(g_{475}-z_{850})}$.
The latter expressions are best-fit relations obtained from our data.
To determine $\alpha$,
we fit via maximum likelihood a function of the 
form $\phi \sim \Gamma_\alpha \hat{\Phi}(m)$ to the observed distribution,
where, as before, $\hat{\Phi}$ is a normal kernel
density estimate of the GCs magnitude distribution
and the behaviour of $\rho_0$ and $r_c^2$ as a function
of magnitude has been obtained with smoothing splines.
While formally it is more sound to include the color factor, we note
that it has a negligible effect on the derived $\alpha$ since
there is no color-luminosity correlation among M87 GCs. 
The best-fit value is $\alpha=1.08\pm0.12$, the resulting
$\phi$ (the solid curve in Figure~\ref{fig:pred_gamma_mag})
shows very good agreement with the data .

To summarize, we have shown that the encounter-rate based quantity
$\Gamma$ is an important factor in determining
$p_X$. Moreover, we have shown that $\Gamma$, and 
the more general parameter $\Gamma_\alpha$,
can account quantitatively for the observed magnitude
distribution of GCs containing LMXBs. Whether one wishes
to assign a more fundamental role  to $\Gamma$ or 
$\Gamma_\alpha$, rather than luminosity, becomes 
to some extent a matter of taste. 
However, we believe that it is more
correct to view $p_X$ as being dependent on 
encounter-rate based quantities, as there are strong theoretical 
arguments to support this interpretation. This is not true of 
the alternative view that $p_X$ depends fundamentally on
cluster luminosity (i.e., more stars do not necessarily imply a 
more favorable environment for the production of compact binaries). 

\subsection{Implications of the Derived Form for $p_X$}
\label{sec:implications}

Combining the results of the previous sections,
our  estimate for $p_X$ is
\begin{equation}
\begin{array}{rcl}
p_X & \sim & \Gamma_{\alpha = 1.08} 10^{\beta(g_{475}-z_{850})} \\ 
        & \sim & \rho_0^{1.08}r_c^2 10^{0.87(g_{475}-z_{850})} \\
        & \sim & \rho_0^{1.08}r_c^2 (Z/Z_{\odot})^{0.33}.\\
\end{array}
\label{eq15}
\end{equation}
Our goal in this section is to use this empirical relation to: (1)
test the validity of the various physical mechanisms which have been 
proposed for the production of LMXBs in dense stellar environments;
and (2) understand the origin of the observed metallicity dependence.

There have been some theoretical suggestions as to why Galactic
LMXBs might be more common in metal-rich GCs.
For instance, Grindlay (1987) noted that, if the IMF depends on
metallicity in such a way as to get flatter with increasing
metallicity, then this would provide a larger population of
massive stars (the progenitors of neutron stars in the LMXBs).
A second, rather different, possibility is the suggestion by
Bellazzini et~al. (1995) that stars of higher metallicity will
have larger radii and higher masses, thereby leading to 
enhancements in the tidal capture rates in metal-rich 
environments. Bellazzini et~al. (1995) also note that such
stars will more easily fill their Roche Lobes, further 
enhancing the number of LMXBs in metal-rich GCs.
Recently, Maccarone et~al. (2004) have suggested that 
irradiation induced winds on the donor star can explain the
observed trend with metallicity.
We now examine these suggestions in detail.

\subsubsection{Dependence of the Number of Compact Stars per Unit 
Initial Mass on Metallicity}
\label{sec:depimf}

Let us denote the number of neutron stars produced per unit initial
mass by $\nu$, the fraction of these neutron stars that are
retained by the cluster by $f$, and the total encounter rate
(including both tidal captures and binary exchanges) by $\Gamma_t$.
Then
\begin{equation}
\Gamma_t \propto f\nu \rho_0^{1.5}r_c^2 (\sigma_{2}+\sigma_{3})T
\label{eq16}
\end{equation}
where $\rho_0$ is the central mass density, 
$r_c$ is the core radius, 
$T$ is the timescale over which the process lasts, and 
$\sigma_{2}$ and $\sigma_{3}$ are the respective cross sections
for tidal capture (two-body) and binary exchange
(three-body) interactions (Johnston et~al. 1992).
Note that the cross-section for tidal capture
scales with the radius, $R$, of the capturing star, while
the binary exchange cross section scales with the semi-major
axis, $a$, of the binary: $i.e.$, 
$\sigma_{2} \propto R$ and $\sigma_{3} \propto a$.

Let us first consider possible metallicity-dependent variations
in the fraction of neutron stars that are retained in GCs. 
Johnston et~al. (1992) show that, for Galactic GCs, $f$ varies
by a factor of $3 \lae f \lae 5$, with the precise value depending
on the assumed velocity distribution of newborn neutron stars. 
For the same sample of GCs, however, 
$\rho_0$ varies by more than two orders of magnitude.
Therefore, even though more massive GCs should certainly
retain a greater fraction of their neutron stars, the net effect
on $\Gamma_t$ is much smaller than that coming from the
increase in the central density $\rho_0$, and we thus neglect it. 
The inclusion of this effect would \textit{steepen} the
dependence of $\Gamma_t$ on $\rho_0$. Note also that for
GCs metallicity does not correlate with mass ($e.g.$,
McLaughlin \& Pudritz 1996), so we expect  $f$ to be independent 
of metallicity. 

We follow usual practice in assuming that $T$ is the same for
the metal-rich and metal-poor subpopulations. Since the
two subpopulations in M87 are observed to be roughly coeval (Cohen,
Blakeslee \& Ryzhov 1998; Jord\'an et~al. 2002; Kissler-Patig
et~al. 2002), this assumption
is reasonable, at least on average, although perhaps questionable
for individual clusters which may have suffered substantial 
dynamical evolution.

Discounting a strong metallicity dependence in $\sigma_2$ or $\sigma_3$ 
(see \S~\ref{sec:radmet}), we are left with the result that 
the best candidate in $\Gamma_t$  to contain the metallicity dependence
of $p_X$ is $\nu$. Thus, we find that the observations point
to a {\it variation in the number of neutron stars formed per unit mass
with GC metallicity}, in the sense that more metal-rich GCs produce
more neutron stars per unit initial mass.
Such a variation could be a consequence
of variations in the IMF, as proposed by Grindlay (1987). But 
even for similar IMFs, metallicity will have important effect
on stellar evolution which will affect the number of 
neutron stars and black holes per unit mass.
For instance, mass loss is thought to be greater for metal-rich
stars and this will have a direct influence in the 
post-main-sequence evolution of high mass stars 
(Heger et~al. 2003).
Thus, the metallicity dependence in $\nu$ can plausibly arise
through more than one process. Below we investigate on 
the possibility of IMF variations
as the cause for the metallicity dependence, but it should be stressed
that the result that $\nu$ should be higher for GCs of higher metallicity
is independent of the IMF being the cause of this dependence.

In describing the IMF we will use a power law description,
with the number of stars with masses between $m$ and $m+dm$ 
given by $N(m) \propto m^{-x}dm$ (where Salpeter is $x=2.35$).
This form is
believed to be a good description of the Galactic IMF 
only for stars with masses $M \gae 1M_{\odot}$;
for lower masses a lognormal distribution appears to be a better
description of the IMF (Chabrier 2003).
Nevertheless, we use the power-law 
form for simplicity and to allow direct comparison with 
previous work. Our main focus below is the effect of IMF
variations in the relative number massive stars that end
their evolution as neutron stars. Using a power law description
this translates into a change in the IMF slope (assuming that all stars
more massive than a certain value end up as neutron stars regardless
of other factors).

If variations in $\nu$ alone are responsible for the observed
scaling of $p_X$ with metallicity through IMF variations, then we can determine the 
behaviour that is required to produce the observed $p_X$.
If all stars with $m > m_{NS}$ evolve to form neutron stars, 
then assuming that the minimum and maximum stellar masses 
are $m_l$ and $m_u$, respectively, the number of
neutron stars per unit mass, $\nu$, is given by
(for $x\neq 1,2$)
\begin{eqnarray}
\nu (x)=\frac{x-2}{x-1}\bigg[\frac{m_u^{1-x}-m_{NS}^{1-x}}{m_u^{2-x}-m_l^{2-x}}\bigg].
\label{eq17}
\end{eqnarray}
We parameterize the dependence of the IMF on metallicity
by assuming $x$ depends linearly on [Fe/H], so
$dx/d\mbox{[Fe/H]}\equiv A$. We restrict
the metallicity range to $-2<\mbox{[Fe/H]}<0$, which includes the
vast majority of M87 GCs (Cohen et~al. 1998). We
assign $x=-2.35$ to [Fe/H]$=-1.0$, assume $m_l=0.08\,M_{\odot}$, 
$m_u=100\,M_{\odot}$, $m_{NS}=8\,M_{\odot}$, and 
then find $A$ by minimizing the quantity 
\begin{eqnarray}
Q = \int_{-2}^{0}(C\nu(\zeta,A)-D10^{0.33\zeta})^2\,d\zeta,
\label{eq18}
\end{eqnarray}
where $C$ and $D$ are normalization constants.
The result of the minimization gives $A=-0.3\pm0.13$~dex$^{-1}$,
where the quoted uncertainty corresponds to the uncertainty
in the metallicity dependence of $p_X$. Thus, the inferred
metallicity dependence is fairly weak.
This result is in good agreement with the analysis of LMXBs in
the Galactic and M31 GC systems presented by Bellazzini
et~al. (1995), who find that $A \sim -0.4$~dex$^{-1}$
is necessary to be consistent with the ratio of LMXBs
in metal-rich and metal-poor clusters. 

In their multivariate analysis of the Galactic GC system,
Djorgovski et~al. (1993) found that the \textit{intrinsic}
dependence of $x$ on metallicity among Galactic GCs ---
after removing the contribution of other important
factors such as galactocentric distance, $R_{GC}$, and
the distance from the Galactic plane, $Z_{GP}$ --- is
$A \approx -0.5$~dex$^{-1}$. Strictly speaking, these scaling
relations are based on the present-day mass function in
GCs, but it is reasonable to assume that this value of
$A$ reflects the dependence of the {\it initial}
mass function on metallicity, as dynamical effects like
cluster evaporation might be implicitly accounted for
by the dependence on $R_{gc}$ and $Z_{gp}$ (Stiavelli et al.
1991).\footnote{The mass function in globular clusters will
undoubtedly change due to dynamical effects,
but as the stellar evolution time scales for $M>8M_{\odot}$
are shorter than dynamical evolution timescales, $\nu$
should be sensitive mainly to the {\it initial} mass function.}
In any event, it is remarkable that our estimate for the
required metallicity dependence of IMF slope is in 
good agreement with that found in the Galaxy. Thus, a difference
in the IMF  between the 
chemically distinct GC subpopulations remains a viable
explanation for the observed metallicity dependence of $p_X$.

\subsubsection{Dependence of Stellar Radii on Metallicity}
\label{sec:radmet}

Bellazzini et~al. (1995) argue that there is a second factor that
can, in principle, increase $p_X$ for metal rich GCs. 
According to these investigators, stars in high-metallicity GCs
will have larger radii and masses than those in metal-poor GCs,
leading to an enhancement in their tidal capture cross sections.
To estimate the magnitude of the effect, Bellazzini et~al. use
the expression for the tidal capture rate, $\Gamma^{TC}$, given
in Lee \& Ostriker (1986):
\begin{equation}
\Gamma^{TC} \propto R^{2-\tau}M^{\tau}N.
\label{eq19}
\end{equation}
\noindent Here $R$ is the radius of the capturing star, $M$ is its mass,
$N$ the total number of such stars, and the power-law index, $\tau = 1.07$,
is appropriate for our case (Lee \& Ostriker 1986). Using the stellar
evolution models of Vandenberg \& Bell (1985), Bellazzini et~al. (1995) 
find a ratio of capture rates for metal-rich and metal-poor stars of 
$\eta \equiv \Gamma^{TC}_{MR} / \Gamma^{TC}_{MP} \sim 2.2$. The enhancement
is therefore comparable to the factor of three difference in $p_X$
which we find for the two GC subpopulations in M87.

However, the precise route to this ratio is not spelled out, and 
we were unable to reproduce their result.
To estimate $\eta$, we
use isochrones of the Bergbush \& Vandenberg (1992), adopt [Fe/H]$=-0.47$
and  [Fe/H]=$-1.82$ for the metal-rich and metal-poor GC subpopulations,
respectively, and assume an age of $13$ Gyr for both subpopulations.
From equation~\ref{eq19}, we find $\eta \sim 2$ if we compare two stars
at the tip of the red giant branch, but this hardly constitutes a
representative estimate for the whole population. 
The latter comparison might be misleading though, as the changes
in the structure of a star as it ascends the red giant branch make
the application of the cross section of Lee \& Ostriker (1986) dubious.
McMillan, Taam \& McDermott (1990) find that the critical impact parameter 
in units of the star radius is smaller for red giant stars than 
main sequence stars, and that it decreases as a star evolves through 
the red giant branch.
In what follows we will neglect this effect, but note
that inclusion of it would further reduce the magnitude
of the effect advocated by Bellazini et~al. (1995).

A representative
estimate of $\Gamma^{TC}$ may be obtained as follows. Assuming that
the IMF is described by a power law, $N(m)=m^{-x}dm$, we set
\begin{equation}
\Gamma^{TC} = \frac{\sum_i R_i^{2-\tau}M_i^\tau m_i^{-x}\Delta m_i}{\sum_i m_i^{-x}\Delta m_i}
\label{eq20}
\end{equation}
\noindent where $i$ runs over all tabulated masses. We thus weight
the tidal capture rates by the expected number of stars at each
mass. Using this approach, we find $\eta \sim 1.1$, with a modest
dependence on the assumed value of $x$. Thus, given similar IMFs,
this effect enhances the tidal capture probability of the metal-rich
GCs by a negligible amount.

A difference in IMF slope between the two GC subpopulations 
(see \S\ref{sec:depimf}) has some effect on the estimated ratio.
In this case, a correction factor ---
similar in form to the one in equation~\ref{eq17}
with $m_{NS}$ replaced by $m_l$, and $m_u$ and $m_l$ set to the
maximum and minimum masses in the isochrones ---
must be included to account for the fact that the number of stars
per unit mass depends on $x$. Denoting such factor by $n(x)$, the
ratio is then
\begin{equation}
\eta = \frac{n(x_{MR})\Gamma^{TC}_{MR}(x_{MR})}{n(x_{MP})\Gamma^{TC}_{MP}(x_{MP})}.
\label{eq21}
\end{equation}
Taking $x=2.35$ for the metal-poor population and $x=1.7$ for the
metal-rich (appropriate for the metallicity dependence of IMF slope in
Galactic GCs, according to Djorgovski et~al. 1993), we 
find $\eta \sim 1.3$. If instead we set $x=1.35$ for
the metal-rich IMF, then the ratio increases to $\eta \sim 1.5$.
Thus, even under rather extreme assumptions for the metallicity 
dependence of IMF slope, the enhancement is insufficient
to explain the observed factor of three difference in $p_X$. 
And, in any event, such a difference in $x$ would result in
a much larger enhancement in $p_X$ through the increase in
the relative numbers of neutron star progenitors (i.e., see 
\S\ref{sec:depimf}). 

To summarize, we suggest that if the form of $p_X$ is
determined solely by dynamical processes, an increase in relative number of
high-mass stars forming in metal-rich environments remains the
only viable explanation for the observed metallicity dependence 
of $p_X$, although a full treatment of the problem should take 
into account the modest variations in tidal capture rates expected
for stars of differing metallicity.

\subsubsection{Irradiation Induced Stellar Winds}
\label{sec:int}

The discussion so far has assumed that $Z$ has no effect
on intrinsic properties of LMXBs, such as their typical lifetimes
and luminosities, which might also produce the observed 
dependence of $p_X$ on metallicity. Recently, Maccarone et~al. (2004) proposed
that irradiation induced stellar winds can explain the metallicity dependence
of $p_X$. The basic mechanism is that radiation induced winds would be
stronger in metal-poor donor stars due to less efficient metal
line cooling and this would speed up the evolution of LMXBs
in metal-poor clusters, leading to the observed trend assuming
other processes such as the ones depicted above are not effective. 
They note that this mechanism may also explain the harder spectra 
observed in metal-poor Galactic LMXBs. 

Even though they do not provide a scaling relation that can be contrasted
directly with the form we determine for $p_X$, they argue
that the ratio of LMXBs for metal rich and metal-poor GCs
(of metallicities $Z_r$ and $Z_p$ respectively) will scale
roughly as $(Z_r/Z_p)^{0.3-0.4}$. This exponent in this scaling
is very similar to the one we derive for $p_X$, and thus the observed
form of $p_X$ is certainly consistent with this scenario.

\subsubsection{Disruption and Hardening of Binaries?}
\label{sec:disrupt}

In terms of $\Gamma$, the dependence of $p_X$ can be written
$p_X \sim \Gamma \rho_0^{-0.42} (Z/Z_{\odot})^{0.33}$. It is
worth reiterating the findings of \S\ref{sec:enc}, namely, that
$\alpha \sim 1.1$, which is reflected in the factor
$\rho_0^{-0.42}$ above.
This result is in agreement with the findings of 
Pooley et~al. (2003) find that the number of close X-ray
binaries in Galactic GCs goes as $N\propto\Gamma^{0.74\pm0.36}$,
which translates into $\alpha\sim1.2$ assuming 
$\sigma\propto\rho_0^{1.5}r_c$. Furthermore, the results are
consistent with the findings of
Johnston et~al. (1992) and Johnston \& Verbunt (1996), which are
not based on LMXBs, but rather on the statistics of
pulsars and low-luminosity X-ray sources in Galactic GCs.
All in all, these findings
point to a scaling of $p_X$ which is shallower than implied
by the value of $\alpha = 1.5$ in equation~\ref{eq11}.

This reduction of $\alpha$ from its ``expected" value of $\alpha=1.5$,
has potentially important implications for the formation and evolution 
of LMXBs. For instance, one possible explanation for this weakening
of the dependence of $\Gamma_t$ on $\rho_0$ would be a change in the
relative importance of $\sigma_{2}$ and $\sigma_{3}$ as a function
of GC central density. Specifically, we can write
the factor which includes the cross sections in $\Gamma_t$ (see
equation~\ref{eq16}) as
${\cal C} \equiv \sigma_{2}+\zeta(\rho_0)\sigma_{3}$. The factor
$\zeta(\rho_0)$ could then account for the fact that, in denser 
clusters, stellar
encounters would be more effective in hardening wide binaries,
thereby reducing $\sigma_{3}$. Since the stellar radii do not depend
on GC structural parameters, we assume $\sigma_{2}$ has no dependence 
on them. Then, from our best-fit
form for $p_X$, we have ${\cal C} \propto \rho_0^{-0.4}$, so that
$\zeta(\rho_0) = (\mu\rho_0^{-0.4} - \sigma_{2})/\sigma_{3}$,
where $\mu$ is a constant. This scaling should
serve as a constrain for predictions on the amount of binary
hardening should this be the cause of the observed reduction
on the expected value of $\alpha$. 

Another possibility is that the {\it destruction} of binaries is
responsible for the observed lessening of the dependence of 
$\Gamma_t$ on $\rho_0$. Once formed, binaries
will be destroyed at a rate $\Delta\Gamma$ that satisfies (Verbunt 2002)
\begin{equation}
\Delta\Gamma \propto \rho_0^{0.5}r_c^{-1}.
\label{eq22_a}
\end{equation}
A simple model would then be that $p_X$ is proportional to the ratio
$\Gamma/{\Delta\Gamma} \propto \rho_0r_c^3 \sim \rho_0^{0.8}r_c^2$, which is 
shallower but still marginally consistent
with the observed, best-fit dependence of $p_X$ on central
density.

\section{Summary and Conclusions}

We have carried out the first detailed study of LMXBs in M87, using a
catalog of 174 X-ray sources identified from deep Chandra/ACIS observations.
All but $\sim$ 20 of these sources are expected to be LMXBs residing
in M87. Combining the X-ray catalog with deep ACS imaging in the $g_{475}z_{850}$
bandpasses for the central 11 arcmin$^2$ of the galaxy, we have explored
the connection between GCs and LMXBs. Our analysis is based on the largest
sample of GC-LMXB associations currently available for any galaxy, and
provides a first glimpse into the relation between compact accretors
and their host GCs in M87.

The luminosity function of X-ray sources
is consistent with a single power law having an upper cutoff at 
$L_X \sim 10^{39}$~erg~s$^{-1}$. Our reanalysis of data in the
literature (Kundu et~al. 2002; Sarazin et~al. 2001) reveals this
also to be the case for M49 and NGC~4697; a
similar conclusion was reached by Sivakoff et~al. (2003) in their
analysis of the LMXB luminosity functions in NGC~4365 and NGC~4382.
We conclude that, contrary to some previous suggestions, there
is no convincing evidence for a break in the luminosity function at
$L_X \sim 3\times10^{38}~$erg~s$^{-1}$ (i.e., the 
Eddington luminosity
corresponding to spherical accretion of ionized hydrogen onto the surface of a
$1.4M_{\odot}$ neutron star). Given the sensitivity of this
luminosity to the nature of the accretion process and
the chemical composition of the accreted material, there seems to
be no {\it a priori} reason to expect a sharp break in the luminosity
function; indeed, we show through numerical simulations that the
features identified by some previous researchers as breaks in the
observed luminosity functions might be a consequence of the
distribution of luminosities having an upper bound.
These findings call into question the usefulness the luminosity function
as a distance indicator, and cautions against using the inferred
breaks to draw conclusions about black hole accretors in early-type
galaxies. If present, such black hole accretors are better probed
through studies of the spectral properties of the detected sources 
({\it e.g.} Irwin et~al. 2003). The luminosity distribution of LMXBs
could remain a viable distance indicator if its form proves to be universal.
The power law exponents we found by fitting truncated power laws are 
marginally consistent 
with a mean value of $\langle \gamma \rangle = -1.78\pm0.08$
for the galaxies we considered; further studies
of expanded samples should be able to test for variations in the luminosity
function slope.

In terms of LMXB formation efficiency in GCs, M87 appears similar to other
well-studied early-type galaxies. We find the percentage of GCs which
contain LMXBs to be $f_X = 3.6\pm0.5$\%, perfectly consistent with the 
values of $2 \lae f_X \lae 4$\% found for a wide variety of early-type
galaxies (Sarazin et~al. 2003). The metal-rich GCs in M87 are observed to be
3$\pm$1 times more likely to contain LMXBs than the metal-poor
GCs, consistent with previous findings for M49 (Kundu et~al. 2002).
All in all, these results for LMXBs mirror other apparently ``universal"
properties of GCs, most notably their near-Gaussian luminosity 
function ($e.g.,$ Harris 2001) and apparently constant
formation efficiency ($e.g.$, Blakeslee et~al. 1997; McLaughlin 1999).
Indeed, the constancy of $f_X$, when coupled with the constant
GC formation efficiency per unit baryon mass (Mclaughlin 1999),
implies a constant LMXB formation efficiency per
unit baryon mass in GCs. It would be interesting to investigate
the bahaviour of the {\it total} number of LMXBs per unit
baryon mass. This could have implications for the proposal that 
most LMXBs may form in GCs (White et~al. 2002, Grindlay 1988), 
and that the presently observed
populations of ``field" LMXBs are the result of GC disruption
and/or LMXB ejection via stellar encounters. As noted by
White et~al. (2002), support for this idea is provided
by the observed scaling of the global LMXB X-ray luminosity
to galactic optical luminosity, $L_{X,glob}/L_{opt}$, with
globular cluster specific frequency (van den Bergh \& Harris 1981).
If all LMXBs are formed in GCs, then a constant LMXB
formation efficiency would arise naturally; on the other hand, if
there are separate populations of
field and GC LMXBs, with different origins, it becomes
harder to explain, as the LMXB population arising from field stars
would have to know about the fraction of baryons that are
\textit{not} in the form of stars. In M87, the observed similarity
between the luminosity distributions of field
and GC LMXBs is broadly consistent with a scenario in which most
LMXBs form in GCs.

In agreement with previous findings based on smaller samples of
LMXB-GC associations, we find that both GC metallicity and luminosity
are important factors in determining the presence of LMXBs in GCs
(Kundu et~al. 2002). 
Furthermore, we have been
able to demonstrate that the probability, $p_X$, that a given
GC will contain a LMXB depends sensitively on the parameter
$\Gamma\equiv\rho_0^{1.5}r_c^2$, which is proportional to the
tidal capture and binary-neutron star exchange rates within
the host GC. {\it This constitutes the strongest evidence to date
that these dynamical processes are responsible for the formation
of the bulk of LMXBs in GCs}.

Working from the subsample of GCs which contain LMXBs and which have 
colors and magnitudes measured from our deep ACS images, we have
explored the scaling of $p_X$ with a variety of GC structural
and photometric parameters. We confirm the previously identified
dependence of $p_X$ on GC metallicity and luminosity, but argue
that the observed luminosity dependence 
arises as a result of the enhanced encounter rates for more
luminous clusters
(mainly because the central mass densities of
GCs increase with increasing luminosity; McLaughlin 2000).
Considering the dependence on structural parameters more fundamental,
our preferred expression for $p_X$ is then
\begin{equation}
p_X \propto \Gamma \rho_0^{-0.42\pm0.11}(Z/Z_{\odot})^{0.33\pm0.1}.
\label{eq22_b}
\end{equation}
The metallicity dependence in this scaling relation --- which
translates into the aforementioned
factor of three enhancement in $p_X$ for metal-rich GCs
relative to their metal-poor counterparts --- has in the past
been proposed to be a result of metallicity-dependent 
variations in the IMF (Grindlay 1987), a consequence
of irradiation induced stellar winds (Maccarone et~al. 2004),
or of 
an enhancement in tidal-capture rates due to the larger radii
of metal-rich stars (Bellazzini et~al. 1995).

We critically examine the viability of these mechanisms in light of the
new observation constraints for M87, and find that previous studies
have likely overestimated the importance of the latter
mechanism. Assuming a universal power law IMF, our calculations suggest
a typical enhancement of $\sim$ 10\% for the metal-rich GCs due
to this radius-metallicity dependence, far
smaller than the observed factor-of-three difference. Only by allowing
the IMF to vary between the chemically distinct GC subpopulations
it is possible to produce such enhancements, but even in this
case, the increase in $p_X$ is driven mainly by the increased
number of neutron star progenitors in metal-rich environments.
On the other hand, the dependence of IMF slope $x$ on metallicity
which is needed to account for the observed metallicity dependence
of $p_X$ is found to be $dx/d\mbox{[Fe/H]}=-0.3\pm0.13$~dex$^{-1}$.
A variation of the IMF slope with metallicity produces
the metallicity dependence in $p_X$ by increasing
the number of compact stars per unit initial mass
for metal-rich GCs. The need for an increased
number of compact stars for higher $Z$ is independent 
of the particular form of the IMF.
We conclude that the only viable dynamical means of accounting for the 
observed metallicity dependence of $p_X$ appears to be a {\it relative
enhancement in the number of neutron stars in
metal-rich GCs}. It is possible that intrinsic properties of LMXBs, 
unrelated to dynamical properties of the host GC, 
are affected by $Z$ and these in term can affect the
form of $p_X$. One such mechanism, irradiation induced
winds, has been recently proposed (Maccarone et~al. 2004) and 
it is consistent with the observed form of $p_X$. Further
detailed studies of the proposed mechanism and contrasting
its predictions with observations should shed light on which 
mechanism -- dynamical or intrinsic -- determines the form of $p_X$.

What emerges from our observations is that a simple dynamical picture
--- namely, the capture of neutron stars by single or binary stars 
within GCs, as fully expressed in equation~(\ref{eq16}) --- can 
account {\it quantitatively} for the observed scaling of $p_X$ with
structural parameters and metallicity. A detailed investigation of
the observed behaviour of $p_X$ should be now undertaken, along with
a comparison to the results of
numerical simulations which probe the formation of compact binaries
in dense stellar environments and incorporate realistic stellar
structure and evolution models. It would be particularly useful
to examine the possibility, discussed in \S\ref{sec:disrupt},
that the effective encounter rates are reduced by a factor
${\Delta}\Gamma \sim \rho_0^{-0.4}$. Thus, our findings for the 
LMXB population in M87 may be evidence for the ongoing disruption of binary
stars in dense environments, or a reduction in the 
binary-neutron star exchange rates due
to the hardening of compact binaries via close encounters ($e.g.$,
Hut et~al. 1992). 

To date, studies of the connection between LMXBs and GCs based on
{\it Chandra} observations of external galaxies have been hindered
by the lack of high-quality optical data needed to characterize
their GC systems. The ACS Virgo Cluster Survey 
(C\^ot\'e et~al. 2004) will offer a nearly complete 
census of GCs within the central regions of one
hundred early-type galaxies in the Virgo Cluster. The measurement of
metallicities, luminosities, perhaps most importantly, structural
parameters, for the many thousands of GCs which will be detected in the
course of this survey will allow refinements to the detailed form
of $p_X$, including an exploration of its possible dependence on
galaxy environment.

\acknowledgments

The authors thank Jack Hughes for illuminating discussions and 
Craig Heinke for useful comments. AJ extends
his thanks to the UCSC Department of Astronomy \& Astrophysics, and
especially to Mike Bolte and Jean Brodie, for their hospitality while
this paper was being prepared. PC acknowledges support for this research 
provided by NASA LTSA grant NAG5-11714, funding for Chandra program
CXC03400562, and funding for HST program GO-9401, 
through a grant from the Space Telescope Science Institute
which is operated by the Association of Universities for 
Research in Astronomy, Inc., under NASA contract NAS5-26555.
Additional support for AJ was provided by the National
Science Foundation through a grant from the Association of
Universities for Research in Astronomy, Inc., under NSF cooperative 
agreement AST-9613615, and by Fundaci\'on Andes under project No.C-13442.
DM acknowledges support from NSF grants AST 00-71099 and AST 02-0631
and from NASA grants NAG5-6037 and NAG5-9046.

\clearpage

\LongTables
\begin{deluxetable}{ccccrrll}
\tabletypesize{\small}
\tablecaption{X-ray Point-sources in M87
\label{tab:cat}}
\tablehead{
\colhead{ID} & \colhead{IAU name} &
\colhead{count rate} &
\colhead{$L_X$\tablenotemark{a}} & \colhead{H21\tablenotemark{b}}
 & \colhead{H31\tablenotemark{b}} & \colhead {ACS} & \colhead{opt. counterpart?} \\
\colhead{} & \colhead{} &
\colhead{10$^{-4}$~cts~s$^{-1}$} &
\colhead{(10$^{37}$~erg~s$^{-1}$)} & \colhead{} & \colhead{} & \colhead{} & \colhead{} 
}
\tablewidth{0pt}
\startdata
1 & CXOU J123033.8+122436 &   1.59 &    2.30 & -0.09 & -0.84 &  no & \nodata  \\
2 & CXOU J123034.8+122527 &   2.98 &    4.29 &  0.17 & -0.13 &  no & \nodata  \\
3 & CXOU J123034.8+122215 &  12.67 &   18.44 &  0.68 &  0.67 &  no & \nodata  \\
4 & CXOU J123035.8+122341 &   3.11 &    4.42 &  0.07 & -0.37 &  no & \nodata  \\
5 & CXOU J123036.0+122433 &   5.41 &    7.71 &  0.26 &  0.11 &  no & \nodata  \\
6 & CXOU J123036.2+122532 &  13.29 &   19.11 &  0.19 & -0.04 &  no & \nodata  \\
7 & CXOU J123036.6+122213 &   3.77 &    5.45 & -0.15 & -0.03 &  no & \nodata  \\
8 & CXOU J123036.8+122459 &   2.78 &    3.96 &  0.15 &  0.48 &  no & \nodata  \\
9 & CXOU J123038.9+122304 &   5.31 &    7.49 &  0.00 & -0.52 &  no & \nodata  \\
10 & CXOU J123039.9+122550 &   2.57 &    4.21 & -0.12 & -0.31 &  no & \nodata  \\
11 & CXOU J123040.3+122509 &   6.37 &    8.97 & -0.17 & -0.36 &  no & \nodata  \\
12 & CXOU J123040.9+122320 &   7.89 &   11.14 &  0.17 & -0.14 &  no & \nodata  \\
13 & CXOU J123041.0+122403 &  18.34 &   25.21 &  0.07 & -0.17 &  no & \nodata  \\
14 & CXOU J123041.2+122450 &   4.14 &    5.81 & -0.29 &  0.09 &  no & \nodata  \\
15 & CXOU J123041.2+122308 &   1.77 &    2.51 & -0.45 & -1.00 &  no & \nodata  \\
16 & CXOU J123041.6+122203 &   3.48 &    4.81 & -0.06 &  0.22 &  no & \nodata  \\
17 & CXOU J123041.6+122600 &  10.45 &   14.71 & -0.02 & -0.25 &  no & \nodata  \\
18 & CXOU J123041.7+122439 &  18.44 &   25.78 &  0.09 & -0.09 &  no & \nodata  \\
19 & CXOU J123042.0+122626 &   1.70 &    2.35 & -0.52 & -0.85 &  no & \nodata  \\
20 & CXOU J123042.0+122450 &   2.35 &    3.29 & -0.35 & -0.65 &  no & \nodata  \\
21 & CXOU J123042.3+122156 &   6.66 &    9.23 &  0.03 & -0.05 &  no & \nodata  \\
22 & CXOU J123042.7+122135 &   4.34 &    8.62 & -0.04 & -0.64 &  no & \nodata  \\
23 & CXOU J123043.0+122525 &   3.36 &    4.71 &  0.17 &  0.07 &  no & \nodata  \\
24 & CXOU J123043.1+122502 &   7.31 &   10.20 & -0.13 & -0.04 &  no & \nodata  \\
25 & CXOU J123043.4+122422 &   5.78 &    8.01 &  0.18 & -0.00 &  no & \nodata  \\
26 & CXOU J123043.4+122746 &   2.85 &    4.06 &  0.28 & -0.34 &  no & \nodata  \\
27 & CXOU J123043.5+122346 &   9.84 &   14.35 &  0.12 & -0.10 &  no & \nodata  \\
28 & CXOU J123043.7+122429 &   3.77 &    5.24 &  0.68 &  0.25 &  no & \nodata  \\
29 & CXOU J123043.7+122418 &   3.26 &    4.51 & -0.16 & -0.60 &  no & \nodata  \\
30 & CXOU J123044.0+122307 &   1.97 &    2.74 &  0.47 &  0.21 &  yes &  no  \\
31 & CXOU J123044.1+122456 &   5.70 &    7.88 &  0.06 & -0.22 &  no & \nodata  \\
32 & CXOU J123044.2+122312 &   2.82 &    3.90 & -0.26 & -0.40 &  yes &  globular  \\
33 & CXOU J123044.2+122134 &  14.09 &   19.24 & -0.17 & -0.22 &  no & \nodata  \\
34 & CXOU J123044.2+122209 &  20.72 &   28.49 &  0.09 & -0.06 &  yes &  no  \\
35 & CXOU J123044.5+122254 &   3.94 &    5.47 &  0.62 &  0.05 &  yes &  no  \\
36 & CXOU J123044.5+122450 &   3.97 &    5.47 &  0.65 &  0.02 &  yes &  globular  \\
37 & CXOU J123044.6+122140 &  27.39 &   38.06 &  0.17 & -0.25 &  no & \nodata  \\
38 & CXOU J123044.6+122201 &  23.83 &   32.65 &  0.27 & -0.06 &  yes &  globular  \\
39 & CXOU J123044.6+122328 &   5.20 &    7.39 &  0.28 & -0.10 &  yes &  globular  \\
40 & CXOU J123044.7+122434 &  49.66 &   68.86 & -0.20 & -0.55 &  yes &  non globular  \\
41 & CXOU J123044.9+122404 &   8.81 &   12.26 &  0.44 &  0.10 &  yes &  no  \\
42 & CXOU J123044.9+122436 &   3.24 &    4.49 & -0.44 & -0.38 &  yes &  globular  \\
43 & CXOU J123045.0+122317 &   5.03 &    6.77 &  0.38 &  0.20 &  yes &  no  \\
44 & CXOU J123045.2+122425 &   5.36 &    7.48 &  0.48 &  0.45 &  yes &  globular  \\
45 & CXOU J123045.3+122352 &   1.98 &    2.75 &  0.60 &  0.55 &  yes &  globular  \\
46 & CXOU J123045.4+122702 &   5.33 &    7.33 &  0.02 &  0.13 &  no & \nodata  \\
47 & CXOU J123045.4+122519 &   3.82 &    5.54 &  0.42 & -0.10 &  no & \nodata  \\
48 & CXOU J123045.4+122329 &   2.93 &    4.19 &  0.21 & -0.65 &  yes &  no  \\
49 & CXOU J123045.5+122412 &   3.34 &    4.66 &  0.63 &  0.31 &  yes &  globular  \\
50 & CXOU J123045.5+122450 &   4.44 &    6.11 & -0.17 & -0.86 &  yes &  no  \\
51 & CXOU J123045.7+122409 &   6.95 &    9.71 &  0.30 & -0.02 &  yes &  globular  \\
52 & CXOU J123045.8+122134 &   3.64 &    5.07 &  0.38 & -0.10 &  no & \nodata  \\
53 & CXOU J123045.8+122336 &   3.36 &    5.35 & -0.12 & -0.30 &  yes &  no  \\
54 & CXOU J123045.9+122408 &   4.42 &    6.17 &  0.29 & -0.12 &  yes &  no  \\
55 & CXOU J123045.9+122125 &   4.96 &    7.68 & -0.15 & -0.12 &  no & \nodata  \\
56 & CXOU J123046.2+122328 &  18.40 &   26.27 & -0.03 & -0.34 &  yes &  globular  \\
57 & CXOU J123046.2+122349 &   4.58 &    6.36 & -0.03 &  0.17 &  yes &  globular  \\
58 & CXOU J123046.3+122207 &   3.00 &    4.12 &  0.63 & -0.62 &  yes &  no  \\
59 & CXOU J123046.3+122323 &  14.49 &   20.62 &  0.17 & -0.01 &  yes &  globular  \\
60 & CXOU J123046.3+122432 &   3.96 &    5.50 & -0.80 & -0.13 &  yes &  globular  \\
61 & CXOU J123046.3+122441 &   3.47 &    4.79 & -0.07 & -1.00 &  yes &  globular  \\
62 & CXOU J123046.5+122450 &   6.75 &    9.33 & -0.17 & -0.16 &  yes &  globular  \\
63 & CXOU J123046.6+122223 &   5.42 &    7.48 & -0.25 & -0.37 &  yes &  globular  \\
64 & CXOU J123046.7+122402 &   9.00 &   12.57 &  0.15 & -0.22 &  yes &  see note 1  \\
65 & CXOU J123046.7+122237 &   3.41 &    4.71 & -0.51 & -0.24 &  yes &  globular  \\
66 & CXOU J123046.7+122453 &   4.84 &    6.70 &  0.00 & -0.12 &  yes &  globular  \\
67 & CXOU J123046.7+122150 &   4.98 &    6.82 & -0.09 & -0.51 &  yes &  globular  \\
68 & CXOU J123046.8+122305 &   8.20 &   11.19 & -0.34 & -0.04 &  yes &  no  \\
69 & CXOU J123046.9+122259 &   3.74 &    5.08 &  0.05 & -0.15 &  yes &  no  \\
70 & CXOU J123047.0+122518 &   2.27 &    4.14 & -0.28 & -0.25 &  no & \nodata  \\
71 & CXOU J123047.0+122500 &   4.59 &    6.37 & -0.19 & -0.52 &  yes &  no  \\
72 & CXOU J123047.1+122415 &  99.35 &  139.27 & -0.07 & -0.65 &  yes &  see note 2  \\
73 & CXOU J123047.3+122308 &  14.80 &   20.10 &  0.08 & -0.23 &  yes &  no  \\
74 & CXOU J123047.5+122126 &   9.66 &   13.54 & -0.04 & -0.10 &  no & \nodata  \\
75 & CXOU J123047.5+122621 &   2.99 &    4.06 & -0.26 &  0.16 &  no & \nodata  \\
76 & CXOU J123047.5+122423 &   7.80 &   10.89 & -0.14 & -0.46 &  yes &  globular  \\
77 & CXOU J123047.6+122351 &   9.18 &   12.80 &  0.10 & -0.60 &  yes &  no  \\
78 & CXOU J123047.6+122234 &   4.54 &    6.25 &  0.15 & -0.04 &  yes &  no  \\
79 & CXOU J123047.6+122220 &   7.20 &    9.93 & -0.03 & -0.11 &  yes &  globular  \\
80 & CXOU J123047.7+122334 &  35.07 &   48.31 &  0.14 & -0.03 &  yes &  globular  \\
81 & CXOU J123047.8+122404 &   7.50 &   10.51 &  0.17 & -0.20 &  yes &  globular  \\
82 & CXOU J123047.8+122401 &   4.88 &    6.85 & -0.12 & -0.47 &  yes &  globular  \\
83 & CXOU J123047.9+122618 &   3.54 &    4.80 &  0.24 & -0.10 &  no & \nodata  \\
84 & CXOU J123048.0+122431 &   3.71 &    5.14 & -0.24 & -0.53 &  yes &  globular  \\
85 & CXOU J123048.3+122455 &   4.54 &    6.32 & -0.14 & -0.15 &  yes &  globular  \\
86 & CXOU J123048.3+122415 &   5.51 &    7.72 &  0.35 & -0.07 &  yes &  globular  \\
87 & CXOU J123048.3+122438 &   4.74 &    6.57 &  0.07 & -0.01 &  yes &  globular  \\
88 & CXOU J123048.7+122620 &   4.99 &    6.75 &  0.57 & -0.32 &  no & \nodata  \\
89 & CXOU J123048.7+122517 &   3.69 &    5.98 & -0.06 & -0.31 &  no & \nodata  \\
90 & CXOU J123048.7+122414 &   9.52 &   13.31 &  0.31 &  0.14 &  yes &  globular  \\
91 & CXOU J123048.8+122347 &  16.35 &   22.81 &  0.07 & -0.48 &  yes &  globular  \\
92 & CXOU J123048.8+122313 &  13.20 &   18.66 &  0.17 & -0.48 &  yes &  no  \\
93 & CXOU J123048.9+122343 &  16.69 &   23.27 & -0.09 & -0.74 &  yes &  no  \\
94 & CXOU J123049.0+122405 &   9.61 &   13.48 &  0.17 & -0.53 &  yes &  globular  \\
95 & CXOU J123049.1+122159 &   9.80 &   13.52 & -0.31 & -0.12 &  yes &  no  \\
96 & CXOU J123049.1+122445 &   2.76 &    3.82 &  0.48 &  0.49 &  yes &  globular  \\
97 & CXOU J123049.1+122604 &  93.37 &  126.71 & -0.04 & -0.50 &  no & \nodata  \\
98 & CXOU J123049.1+122308 &   3.71 &    5.87 &  0.32 & -0.12 &  yes &  globular  \\
99 & CXOU J123049.2+122334 &  79.06 &  109.47 & -0.22 & -0.80 &  yes &  globular  \\
100 & CXOU J123049.5+122355 &  12.17 &   17.05 &  0.07 & -0.55 &  yes &  globular  \\
101 & CXOU J123049.6+122333 &  26.42 &   36.62 &  0.11 & -0.33 &  yes &  globular  \\
102 & CXOU J123049.6+122353 &   3.67 &    5.14 &  0.16 & -0.42 &  yes &  globular  \\
103 & CXOU J123049.7+122351 &  15.10 &   21.12 &  0.19 & -0.30 &  yes &  globular  \\
104 & CXOU J123049.8+122402 &  22.62 &   31.70 &  0.22 & -0.32 &  yes &  globular  \\
105 & CXOU J123049.8+122216 &   2.38 &    3.27 & -0.21 & -0.69 &  yes &  no  \\
106 & CXOU J123049.8+122436 &   8.49 &   11.73 & -0.08 & -0.20 &  yes &  no  \\
107 & CXOU J123049.9+122740 &   4.37 &    6.59 &  0.27 &  0.02 &  no & \nodata  \\
108 & CXOU J123050.0+122400 &  17.67 &   24.77 &  0.16 & -0.27 &  yes &  globular  \\
109 & CXOU J123050.1+122251 &  10.98 &   14.97 &  0.04 & -0.45 &  yes &  globular  \\
110 & CXOU J123050.1+122301 &  23.81 &   33.50 &  0.10 & -0.35 &  yes &  globular  \\
111 & CXOU J123050.2+122608 &   4.65 &    6.27 & -0.11 & -0.47 &  no & \nodata  \\
112 & CXOU J123050.3+122128 &   3.32 &    4.59 &  0.58 &  0.01 &  no & \nodata  \\
113 & CXOU J123050.3+122332 &   5.65 &    7.84 &  0.23 & -0.34 &  yes &  globular  \\
114 & CXOU J123050.4+122212 &   5.70 &    7.82 & -0.32 & -0.34 &  yes &  globular  \\
115 & CXOU J123050.5+122356 &  18.38 &   25.75 &  0.17 & -0.46 &  yes &  no  \\
116 & CXOU J123050.5+122435 &   3.11 &    4.29 & -0.79 & -0.64 &  yes &  globular  \\
117 & CXOU J123050.8+122502 &  43.15 &   69.31 &  0.12 & -0.28 &  yes &  no  \\
118 & CXOU J123050.8+122411 &   6.49 &    9.04 & -0.32 & -0.11 &  yes &  no  \\
119 & CXOU J123051.1+122242 &   3.52 &    4.77 &  0.41 & -0.09 &  yes &  globular  \\
120 & CXOU J123051.8+122159 &   5.47 &    7.49 &  0.67 &  0.22 &  yes &  no  \\
121 & CXOU J123051.8+122247 &   2.89 &    3.94 &  0.15 &  0.23 &  yes &  no  \\
122 & CXOU J123051.8+122911 &   1.41 &    2.13 & -0.12 & -1.00 &  no & \nodata  \\
123 & CXOU J123052.5+122533 &   4.90 &    6.75 &  0.56 &  0.24 &  no & \nodata  \\
124 & CXOU J123052.6+122323 &   3.95 &    5.49 &  1.00 &  1.00 &  yes &  no  \\
125 & CXOU J123052.7+122336 &   7.19 &   10.06 &  0.38 &  0.27 &  yes &  globular  \\
126 & CXOU J123052.9+122547 &   7.34 &    9.99 & -0.04 & -0.21 &  no & \nodata  \\
127 & CXOU J123053.0+122244 &   3.97 &    5.40 & -0.17 & -0.88 &  yes &  no  \\
128 & CXOU J123053.0+122535 &   4.27 &    5.84 &  0.25 & -0.28 &  no & \nodata  \\
129 & CXOU J123053.0+122208 &   4.46 &    6.07 & -0.63 & -0.18 &  yes &  no  \\
130 & CXOU J123053.2+122356 &  28.14 &   39.40 &  0.10 & -0.20 &  yes &  globular  \\
131 & CXOU J123053.4+122556 &   3.16 &    4.26 &  0.22 &  0.28 &  no & \nodata  \\
132 & CXOU J123053.6+122237 &   4.01 &    5.44 & -0.42 & -0.07 &  yes &  non globular  \\
133 & CXOU J123053.7+122448 &   2.91 &    3.99 &  0.25 & -0.51 &  yes &  no  \\
134 & CXOU J123053.9+122430 &   4.73 &    6.51 & -0.42 & -0.12 &  yes &  globular  \\
135 & CXOU J123053.9+122544 &   6.59 &    8.96 &  0.25 & -0.44 &  no & \nodata  \\
136 & CXOU J123054.4+122302 &   7.01 &   10.07 &  0.02 &  0.14 &  yes &  globular  \\
137 & CXOU J123054.6+122218 &   1.97 &    2.67 & -0.32 &  0.17 &  yes &  no  \\
138 & CXOU J123054.7+122222 &   7.46 &   10.12 &  0.28 & -0.22 &  yes &  globular  \\
139 & CXOU J123054.8+122231 &   3.42 &    4.65 &  0.23 & -0.46 &  yes &  no  \\
140 & CXOU J123054.9+122247 &   4.22 &    5.98 & -0.04 & -0.46 &  yes &  no  \\
141 & CXOU J123054.9+122538 &  22.17 &   30.24 &  0.07 & -0.12 &  no & \nodata  \\
142 & CXOU J123054.9+122438 &   8.14 &   11.20 &  0.27 &  0.16 &  yes &  globular  \\
143 & CXOU J123055.0+122504 &   2.83 &    4.05 &  0.57 & -0.76 &  yes &  no  \\
144 & CXOU J123055.2+122340 &  11.83 &   16.63 & -0.03 & -0.03 &  yes &  no  \\
145 & CXOU J123055.3+122256 &   3.41 &    5.23 & -1.00 & -0.50 &  yes &  globular  \\
146 & CXOU J123055.4+122314 &   3.81 &    5.31 & -0.34 &  0.01 &  yes &  globular  \\
147 & CXOU J123055.4+122342 &  10.93 &   15.35 &  0.15 &  0.06 &  yes &  globular  \\
148 & CXOU J123055.5+122615 &   2.64 &    3.56 &  0.24 & -0.13 &  no & \nodata  \\
149 & CXOU J123056.2+122526 &   2.49 &    3.40 &  1.00 &  1.00 &  no & \nodata  \\
150 & CXOU J123056.2+122447 &   8.26 &   11.79 &  0.12 & -0.02 &  yes &  globular  \\
151 & CXOU J123056.3+122211 &  11.71 &   15.95 &  0.12 & -0.18 &  yes &  globular  \\
152 & CXOU J123056.4+122448 &   4.65 &    7.39 &  0.02 & -0.26 &  yes &  globular  \\
153 & CXOU J123056.7+122259 &   6.44 &    8.93 &  0.06 & -0.50 &  yes &  no  \\
154 & CXOU J123057.2+122122 &  14.09 &   19.44 &  0.13 & -0.04 &  no & \nodata  \\
155 & CXOU J123057.9+122220 &   3.92 &    5.37 &  0.48 &  0.34 &  no & \nodata  \\
156 & CXOU J123058.0+122844 &   1.57 &    2.31 &  0.24 &  0.13 &  no & \nodata  \\
157 & CXOU J123058.1+122104 &   9.51 &   13.16 &  0.20 & -0.31 &  no & \nodata  \\
158 & CXOU J123058.5+122222 &  10.03 &   13.77 &  0.13 & -0.25 &  no & \nodata  \\
159 & CXOU J123059.0+122259 &   3.47 &    4.83 &  0.13 & -0.59 &  no & \nodata  \\
160 & CXOU J123059.5+122038 &  38.97 &   54.50 & -0.23 & -0.51 &  no & \nodata  \\
161 & CXOU J123059.6+122509 &  25.13 &   34.53 &  0.02 & -0.36 &  no & \nodata  \\
162 & CXOU J123100.1+122232 &   4.58 &    6.58 & -0.16 & -0.51 &  no & \nodata  \\
163 & CXOU J123100.2+122138 &   3.37 &    4.69 & -0.00 & -0.45 &  no & \nodata  \\
164 & CXOU J123100.3+122417 &  14.62 &   20.30 &  0.27 &  0.08 &  no & \nodata  \\
165 & CXOU J123100.6+122021 &   1.30 &    3.12 & -0.42 & -0.37 &  no & \nodata  \\
166 & CXOU J123102.4+122040 &   1.52 &    2.14 &  0.36 &  0.15 &  no & \nodata  \\
167 & CXOU J123102.6+122121 &   2.21 &    3.12 &  0.18 & -0.70 &  no & \nodata  \\
168 & CXOU J123102.6+122411 &   5.06 &    7.05 &  0.04 & -0.52 &  no & \nodata  \\
169 & CXOU J123103.3+122107 &   3.71 &    5.24 & -0.38 & -0.21 &  no & \nodata  \\
170 & CXOU J123103.5+122305 &   2.30 &    3.22 & -0.17 & -1.00 &  no & \nodata  \\
171 & CXOU J123104.1+122335 &   1.71 &    2.39 & -0.20 & -1.00 &  no & \nodata  \\
172 & CXOU J123104.5+122156 &   2.97 &    4.80 &  0.21 &  0.14 &  no & \nodata  \\
173 & CXOU J123105.5+122731 &   2.35 &    3.38 &  0.28 & -0.30 &  no & \nodata  \\
174 & CXOU J123108.9+122618 &   1.03 &    1.53 &  0.27 &  0.27 &  no & \nodata  \\
\enddata
\tablenotetext{a}{$L_X$ values are in the $0.3-10$ keV energy band and are obtained assuming
a power law spectral shape with power law index $\kappa=1.64$ and a Galactic
hydrogen column density $N_H=2.5\times10^{20}$ cm$^{-2}$ (Stark et~al. 1992).}
\tablenotetext{b}{H$21\equiv (M-S)/(M+S)$ and H$31\equiv (H-S)/(H+S)$ where $S$, $M$
and $H$ are the counts in a soft ($0.3-1$ keV), medium ($1-2$ keV), and
hard ($1-10$ keV) energy bands respectively.}
\tablenotetext{1}{Two optical candidates within matching radius.}
\tablenotetext{2}{Elongated X-ray source located within three adjacent optical sources.}
\end{deluxetable}

\clearpage

\begin{deluxetable}{lccc}
\tablecaption{LMXB Luminosity Function Parameters
\label{tab:xlf}}
\tablehead{
\colhead{Parameter} & \multicolumn{3}{c}{Value} \\
\cline{2-4}\\
\colhead{}          & \colhead{All} & \colhead{GCs} & \colhead{Field} \\
}
\tablewidth{0pt}
\startdata
{\it Broken Power Law}    &       & & \\
$\alpha_1$                & $-1.16\pm0.25$ & $ -1.12\pm0.18$& $-1.38\pm0.28$\\
$\alpha_2$                & $-1.86\pm0.54$ & $-4.8\pm1.7$ & $-4.89\pm1.1$ \\
$\log(L_B[$erg~s$^{-1}])$   & $38.30\pm 0.42$& $38.49\pm 0.18$& $38.34\pm 0.15$ \\
\tableline
{\it Truncated Power Law} &    & &    \\
$\gamma$                  & $-2.14\pm0.14$ & $-2.07\pm0.20$ & $-2.36\pm0.33$ \\
\enddata
\end{deluxetable}

\begin{deluxetable}{ccc}
\tablecaption{Model Metallicities and Colors for Globular Clusters\tablenotemark{a}
\label{tab:bc}}
\tablehead{
\colhead{[Fe/H]} & \colhead{$(g_{475}-z_{850})$} & \colhead{$(V-z_{850})$}\\
\colhead{(dex)} & \colhead{(AB mag)} & \colhead{(AB mag)}
}
\tablewidth{0pt}
\startdata
$-$2.3 & 0.851 & 0.493 \\
$-$1.7 & 0.907 & 0.512 \\
$-$0.7 & 1.243 & 0.763 \\
$-$0.4 & 1.425 & 0.901 \\
$+$0.0 & 1.588 & 1.022 \\
$+$0.4 & 1.897 & 1.257 \\
\enddata
\tablenotetext{a}{Obtained from the Bruzual \& Charlot (2003) models
assuming an age of $13$ Gyr}
\end{deluxetable}

\end{document}